\documentclass[twocolumn,showpacs,pre,amsmath,amssymb,floatfix]{revtex4}
\usepackage{graphicx}
\usepackage{dcolumn}
\usepackage{bm}

\begin{document}

\title{Structural dependence of robustness and load tolerance in scale-free networks}

\author{Nobuhiko Oshida}
\author{Sigeo Ihara}
\affiliation{Department of Advanced Interdisciplinary Studies, and Research Center for Advanced Science and Technology, The University of Tokyo, Tokyo 153-8904, Japan}

\date{\today}

\begin{abstract}
The structure of complex networks in previous research has been widely described as scale-free networks generated by the preferential attachment model.
However, the preferential attachment model does not take into account the detailed topological property observed in real networks.
Here we propose the models of scale-free networks which can reproduce the topological assortativity of real networks.
With an identical degree distribution and network size, we study the structural robustness and fragility as well as the dynamic change of load intensity when nodes are successively removed under random and various intentional attack strategies.
We find that disassortative networks are structurally robust against random attacks and highly load-tolerant, while assortative networks are most resistant to intentional attacks yet significantly fragile against random attacks.
\end{abstract}


\maketitle

\section{Introduction}
Many complex systems such as the Internet, electric circuits, and biological cells can retain their functions despite strong perturbations.
The robustness of these systems is often attributed to underlying heterogeneous network structure: only a few nodes called hubs have a large number of edges, but the majority of nodes have a few edges.
Those structures are called scale-free networks, where the degree distribution $P(k)$, the probability a node has $k$ edges, decays as a power-law $P(k) \propto k^{-\gamma}$ \cite{albert2002, dorogovtsev2003}.

The existing studies of network robustness mainly focus on preservation of connectivity against successive attacks of node removal.
Callaway {\it et al.} \cite{callaway2000} and Cohen {\it et al.} \cite{cohen2000, cohen2001} have analytically calculated the critical thresholds at which networks break apart under both random and intentional attacks based on percolation theory.
However, their methods are applicable when cycles, closed paths with no other repeated nodes than the starting and ending nodes, can be ignored.
Most of the real networks are highly clustered and contain many cycles; therefore, their critical thresholds have usually been derived by computational simulations \cite{xulvi-brunet2003, serrano2006-l, serrano2006-2}.
The current consensus on the connectivity robustness in complex networks is that, as compared to random networks, scale-free networks are robust against random failures, yet fragile against intentional attacks on hubs \cite{albert2000, gallos2005}.
It has been also reported that the attack strategy based on betweenness centrality much harms network connectivity \cite{holme2002}.
These results are obtained with scale-free networks created through the preferential attachment rule \cite{barabasi1999}, but their inner structures do not reflect degree-degree correlations observed in real networks \cite{newman2002}.
As shown in our recent paper \cite{oshida2006}, the traffic efficiency on scale-free networks significantly varies depending on their degree-degree correlations.
Here we are interested in the effect of inner structures on connectivity robustness in scale-free networks.

Our interest also lies in the dynamic change of load intensity caused by traffic flows on scale-free networks.
To maintain the stability against heavy load attributed to node removal, network structures should be highly load-tolerant.
Although it is known that the excess load tends to cause the lethal avalanches of node breakdowns \cite{moreno2002, motter2002} and traffic congestion \cite{zhao2005, lee2005, wu2006}, little attention has been paid to the effect of inner structures on load intensity when networks are successively attacked.
We thus consider the load tolerance in addition to connectivity robustness to evaluate the reliability of networks.

In this article, we introduce the models of scale-free networks which take into account the topological assortativity of real networks.
With an identical degree distribution and network size, we investigate the connectivity robustness as well as the load tolerance when nodes are successively removed under random and various intentional attack strategies.
From these analyses, we study the relationship between inner structures and robustness in scale-free networks.

\section{Methods}
\subsection{Network models}
We introduce four models of scale-free network which generate different topological characteristics.
First we construct a scale-free network with the preferential attachment (PA) rule \cite{barabasi1999}, which is a basic growth algorithm to produce a network with a power-law degree distribution: starting with a small complete graph, each new node with $m$ edges is added and connected to different old nodes with probabilities proportional to their degrees.
To analyze the percolation properties of realistic networks, the minimum degree $k_{\min}(=m)$ is two since we cannot make cycle structures in the case of $m<2$.
In addition, we employ the uncorrelated null (UN) model \cite{maslov2002} which generates a network by rewiring all edges randomly using the PA network.
Both of the PA and UN networks have non-assortative structures.
While the PA networks contain age correlations because of growing scheme of network construction, the UN model can destroy all the hidden correlations in the PA network.

We propose two other models which create scale-free networks reproducing the inner structures of real networks.
Our models change only assortative level of the network with fixing the degree distribution $P(k)$ of the PA model.
One of the models generates disassortative networks where high-degree nodes preferably connect to low-degree nodes (HL).
The other model generates the network with assortative topology where high-degree nodes preferably connect to other high-degree nodes (HH).
First, based on the degree distribution $P(k)$, all nodes are labeled and arranged in a queue as $v_1, v_2, \cdots, v_N$ in descending order of their degrees.
The HL network is produced by the algorithm iterating the following procedures:
selecting node $v_i$ with the highest degree $k_i$ from the head of the queue, linking edges between $v_i$ and $k_i$ nodes picked up randomly from the queue ($v_{i+1}, \cdots, v_{N}$), and removing nodes fully connected from the queue.
Similarly, the HH network is produced by the algorism of the HL model with replacing the second procedure:
linking edges between $v_i$ and $k_i$ nodes picked up randomly from the queue depending on their weights $w_{i+1}, \cdots, w_{N}$ ($w_j = k_j^n, n > 1$).
In this study, we set $n=3$ so that the absolute values of assortativity of both the HL and HH networks become nearly equivalent (see Table \ref{tab:tab1}).
The topology produced by the HL model is seen in biological networks and the Internet \cite{newman2002, maslov2002, vazquez2002, newman2003, rual2005, serrano2006-1, guimera2007}, whereas the topology produced by the HH model is seen in social networks \cite{newman2002, newman2003, newman2003-2}.

For reference, we also construct the random network based on the Erd\H{o}s-R\'enyi (ER) model \cite{erdos1959, dorogovtsev2003} whose minimum degree is also two.
All the networks in this study have 1000 nodes and 2000 edges.
In case a slight topological difference may change the results of analysis, the results shown in this paper are averaged by 1000 trials.

\begin{figure*}[tbp]
	\begin{center}
		\begin{tabular}{lll}
			\includegraphics[scale=1.0, clip]{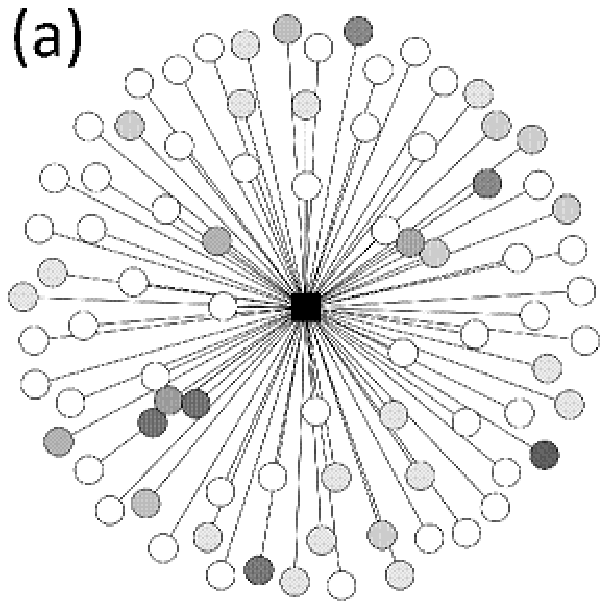} & \ \ \ \includegraphics[scale=1.0, clip]{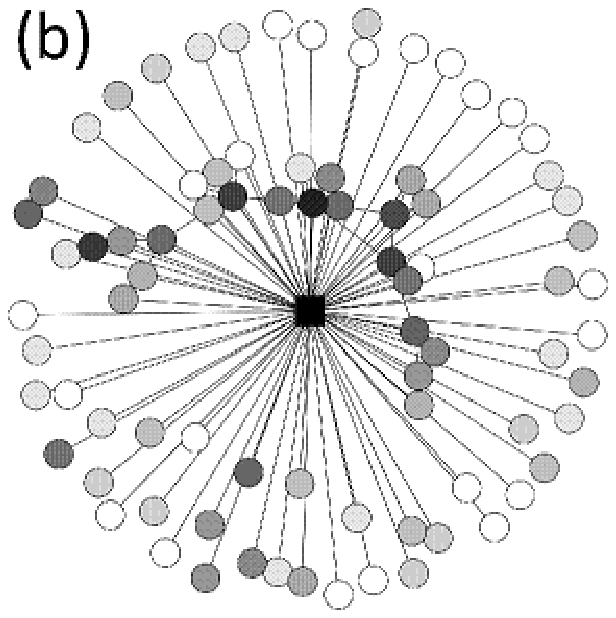}\\ \\ \\
			\includegraphics[scale=1.0, clip]{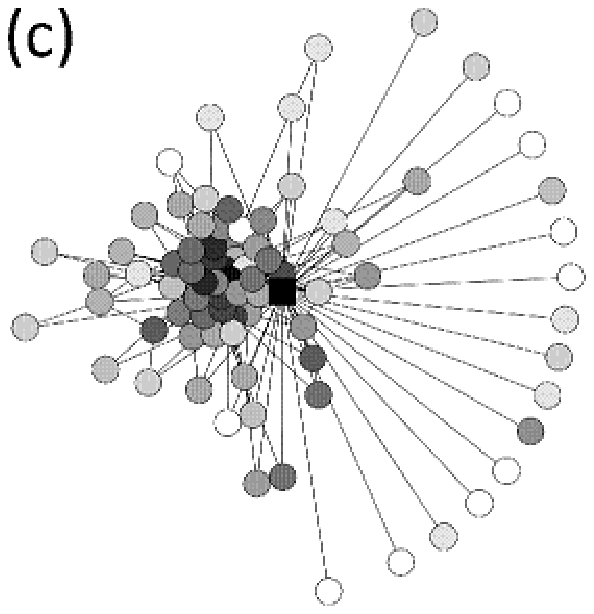} & \ \ \ \includegraphics[scale=1.0, clip]{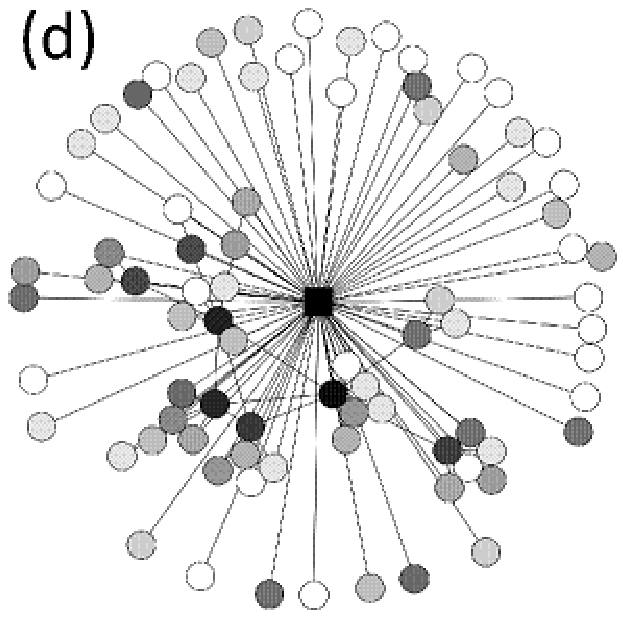} & \ \ \ \includegraphics[scale=0.5]{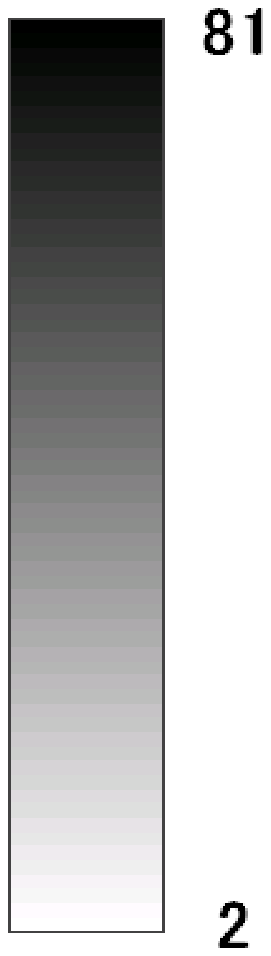}
		\end{tabular}
	  	\caption{Examples of neighbor connections adjacent to a hub with the most edges in the (a) HL, (b) UN, (c) HH and (d) PA networks. Each hub is represented as a square. Continuous color map reflects the log-scaled degree $k$ of a node ($2 \leq k \leq 81$).}
 		\label{fig:fig1}
	\end{center}
\end{figure*}

Although these four types of scale-free networks have the identical degree distribution, their topological properties are quite different.
Fig.\ref{fig:fig1} shows the local connectivity of the node with the highest degree generated by each network model.
In the HL network, the nodes adjacent to hubs have low degree.
This means that there are few direct edges between hubs.
In contrast, the hub in the HH network connects to other high-degree nodes.
Both of the PA and UN networks display the intermediate characteristics between the HL and HH networks.

\subsection{Topological indices}
We quantitatively evaluate the structural properties of those networks with some topological indices.
The degree-degree correlation expresses the topological tendency between the degree of a node and that of the nearest neighbors, and it is evaluated with assortativity coefficient \cite{newman2002, newman2003} expressed as
\begin{eqnarray}
r=\frac{M^{-1}\sum_{i}j_{i}k_{i}-[M^{-1}\sum_{i}\frac{1}{2}(j_{i}+k_{i})]^{2}}{M^{-1}\sum{\frac{1}{2}(j_{i}^{2}+k_{i}^{2})-[M^{-1}\sum_{i}\frac{1}{2}(j_{i}+k_{i})]^{2}}},
\end{eqnarray}
where $j_{i}$ and $k_{i}$ are the degree of the nodes at the end of the $i$th edge, and $M$ is the number of edges.
The assortativity $r$ ranges between $-1 \leq r \leq 1$ depending on the assortative level of the network.

The average path length ${\overline L}$ of a network is given by
\begin{eqnarray}
{\overline L} = \frac{2\sum_{i<j}L_{ij}}{N(N-1)},
\end{eqnarray}
where $L_{ij}$ is the shortest path length between node $v_i$ and node $v_j$, and $N$ is the number of nodes.

The clustering characteristic of node $v_i$ is evaluated by the clustering coefficient $C_i$ expressed as
\begin{eqnarray}
C_i=\frac{2E_{i}}{k_{i}(k_{i}-1)},
\end{eqnarray}
where $k_{i}$ is the degree of node $v_i$ and $E_{i}$ is the number of edges that exist between these $k_{i}$ nodes.
The average clustering coefficient of the network is given by ${\overline C}=\sum_iC_i/N$.

The  betweenness centrality $B_i$ of node $v_i$ is given by
\begin{eqnarray}
B_i=\sum_{s<t}\frac{\sigma_{st}^i}{\sigma_{st}},
\end{eqnarray}
where $\sigma_{st}$ is the total number of the shortest paths from node $v_s$ to node $v_t$ and $\sigma_{st}^i$ is the number of the shortest paths from $v_s$ to $v_t$ passing through node $v_i$.
Nodes of high betweenness centrality have a key role to shorten the length of many paths between nodes in the network \cite{freeman1977, barthelemy2004}, and tend to have high load intensity \cite{goh2001, oshida2006}.
The average betweenness centrality of the network is given by ${\overline B}=\sum_iB_i/N$

The distribution of betweenness centrality is an important factor to estimate the efficiency of traffic load on the network \cite{zhao2005, oshida2006}, and we thus evaluate the betweenness deviation expressed as
\begin{eqnarray}
\delta_{B}=\sqrt{\frac{\sum_i(B_{i}-\overline{B})^2}{N}}.
\end{eqnarray}
The low betweenness deviation means that load on the network is efficiently distributed, whereas the high betweenness deviation means that load is concentrated in a part of the network.

As shown in Table \ref{tab:tab1}, the HL model shows disassortativity; that is, there are many edges between a hub and a low-degree node.
This disassortative characteristic of the HL network slightly lengthens average path length and lessens clustering characteristics compared to the PA network.
Furthermore, the betweenness deviation of the HL network is significantly smaller than that of the PA network.
This means that the load on the HL network is efficiently distributed.
Although the HH network has the same power-law degree distribution, it shows high assortativity and has the long average path length.
In addition, the average betweenness centrality of the HH network is quite high.
This indicates that there are many bottlenecks in the HH network in the initial state.
In the PA and UN networks, $r$ is nearly zero, indicating non-assortativity; that is, their structures have no degree-degree correlation.
However, the characteristics of ${\overline L}$, ${\overline B}$, and $\delta_{B}$ in the UN network are largely different from those in the PA network.
These differences indicate that age correlation affects the topological characteristics of network.

\subsection{Attack strategies}
The robustness of networks has been previously analyzed under the strategies of node attack at random or in order of node's degree \cite{albert2000}.
In addition to the degree-based attack, the betweenness-based attack also belongs to an intentional harmful strategy.
The order of these intentional attacks can be determined with the initial distributions of degree or betweenness centrality.
However, as nodes are removed from the network, these distributions change accordingly.
The order of nodes determined by their initial degrees or betweenness centralities does not usually coincide with the recalculated order after the removal of nodes.

Given the variety of attack orders, the intentional strategies are classified depending on whether they are based on degree or betweenness centrality and whether the order of node attack is determined statically or dynamically \cite{holme2002}.
The static degree-based attack (SDA) is the strategy that the target nodes are selected one by one in descending order of their initial degrees.
The dynamic degree-based attack (DDA) is the strategy that the target nodes are dynamically selected by searching a node with the highest degree at every removal step.
In the same way, we can also define the static betweenness-based attack (SBA) and the dynamic betweenness-based attack (DBA) which target the node with the highest betweenness centrality.
In addition to these strategies, we adopt the random attack (RA) that the target nodes are selected randomly.
When a node is removed from networks, all edges connecting the node are also removed.

\section{Results}
\subsection{Connectivity robustness}
\begin{figure*}[tbp]
	\begin{center}
		\begin{tabular}{c}
			\includegraphics[scale=0.45]{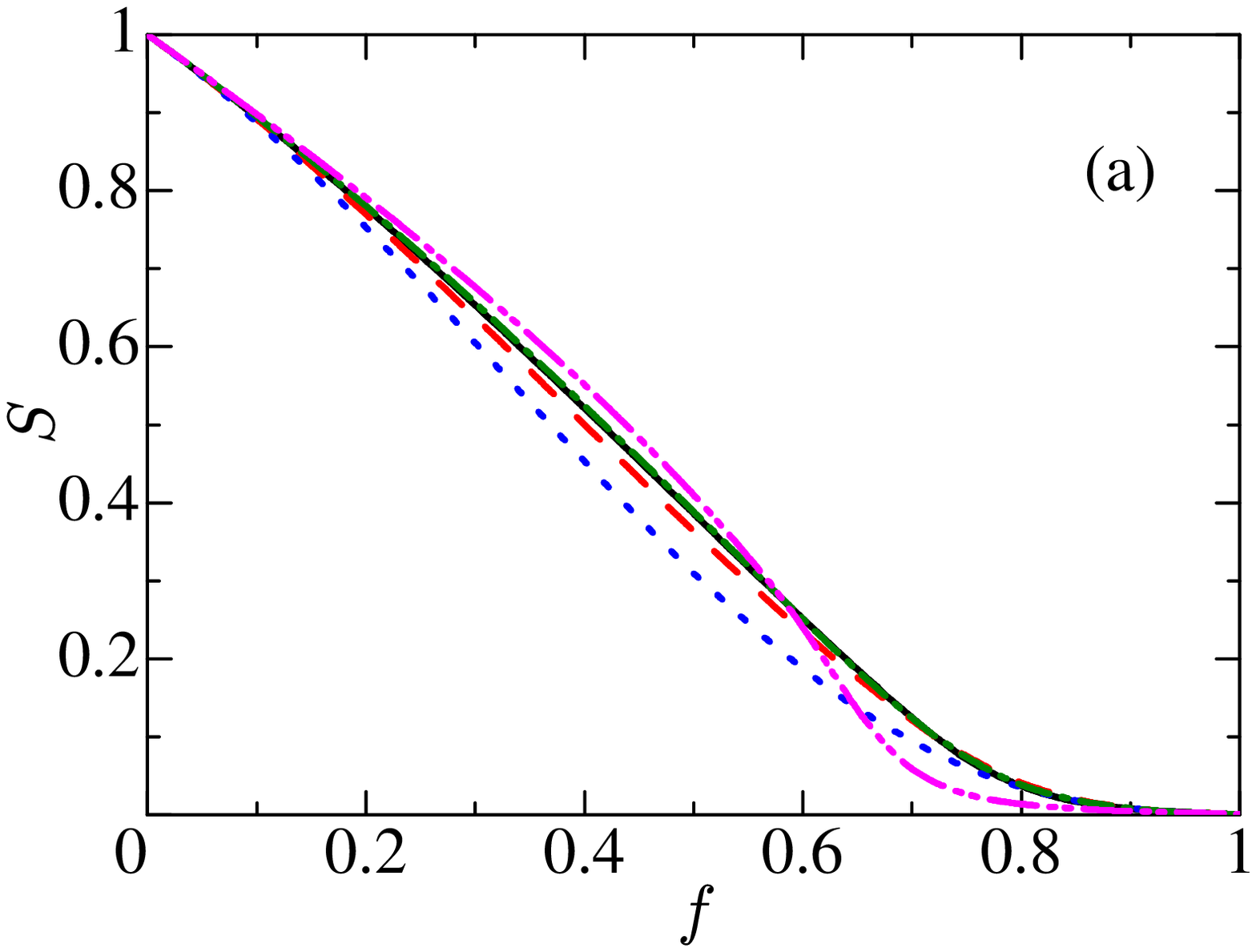}\ \ \ \ \ \ \ \ \
			\includegraphics[scale=0.45]{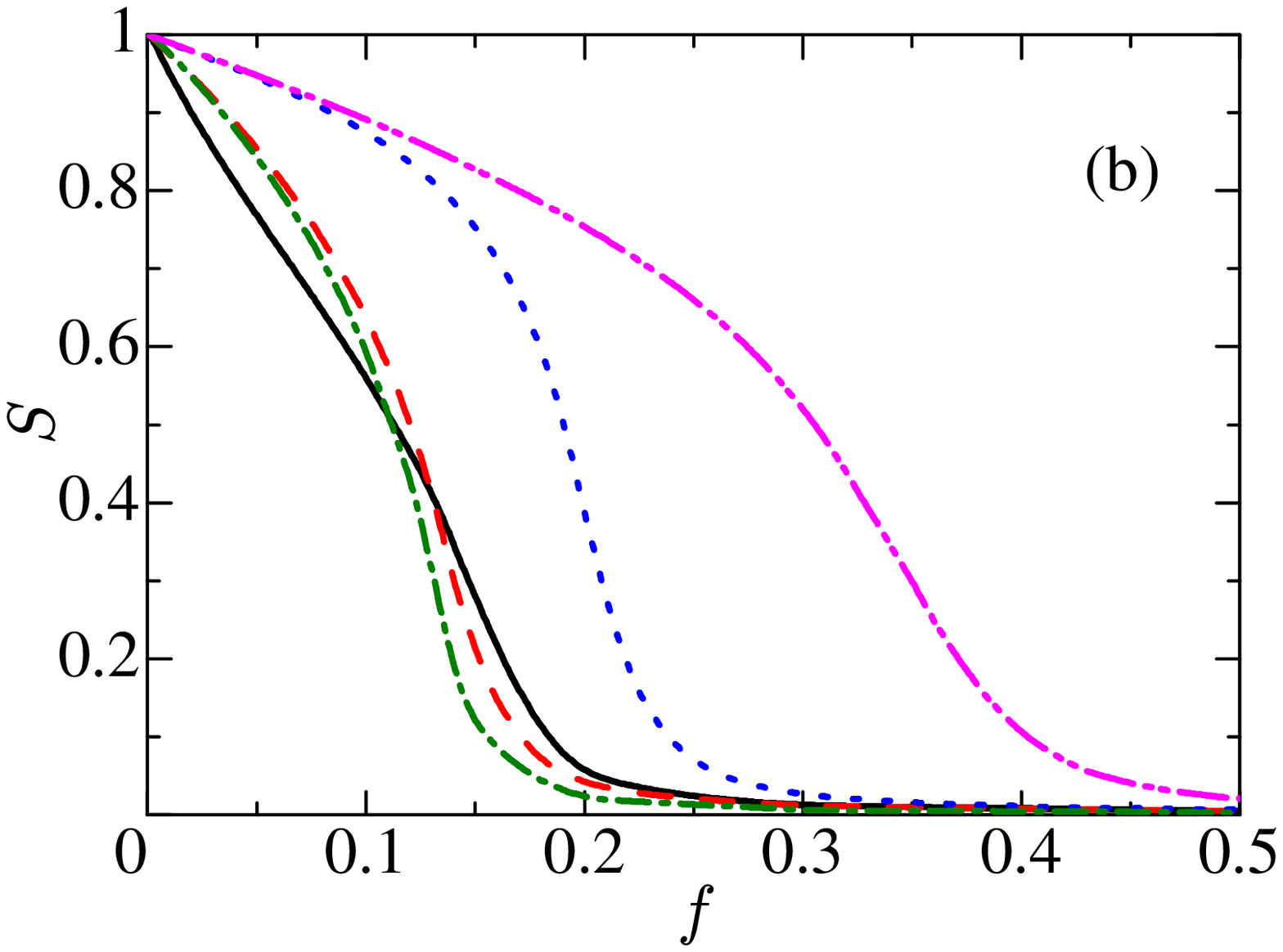}\\ \\
			\includegraphics[scale=0.45]{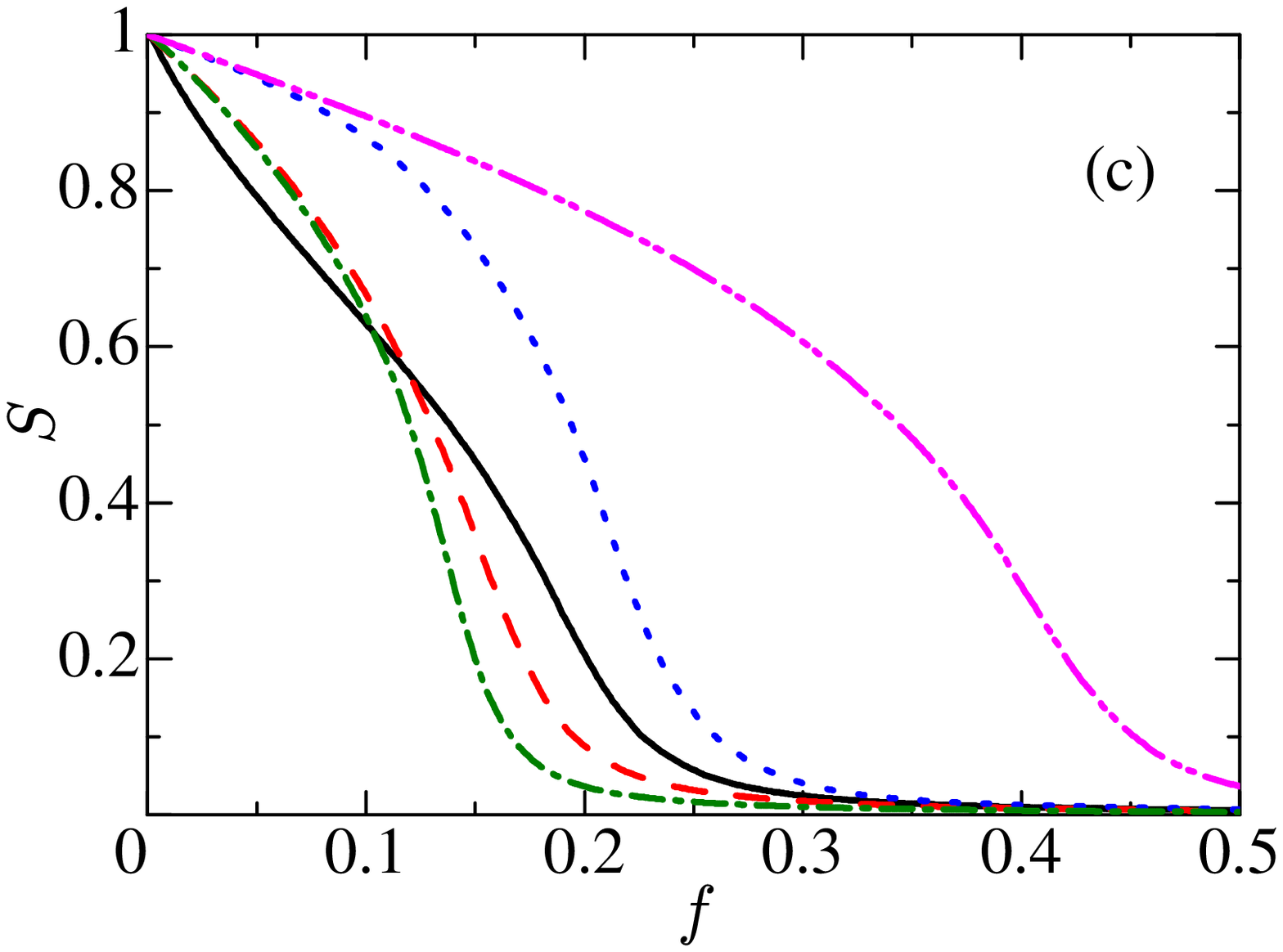}\ \ \ \ \ \ \ \ \
			\includegraphics[scale=0.45]{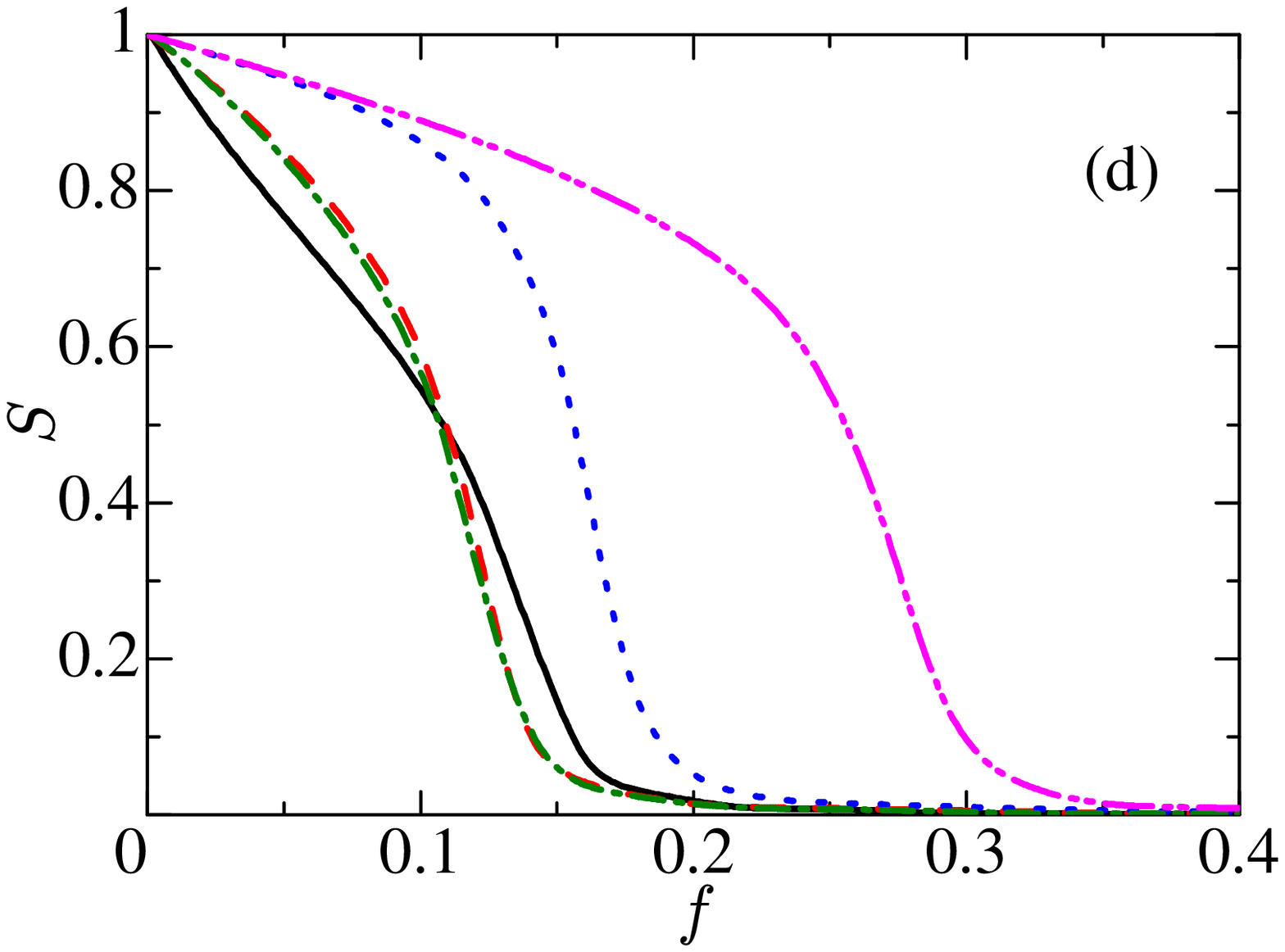}\\ \\
			\includegraphics[scale=0.45]{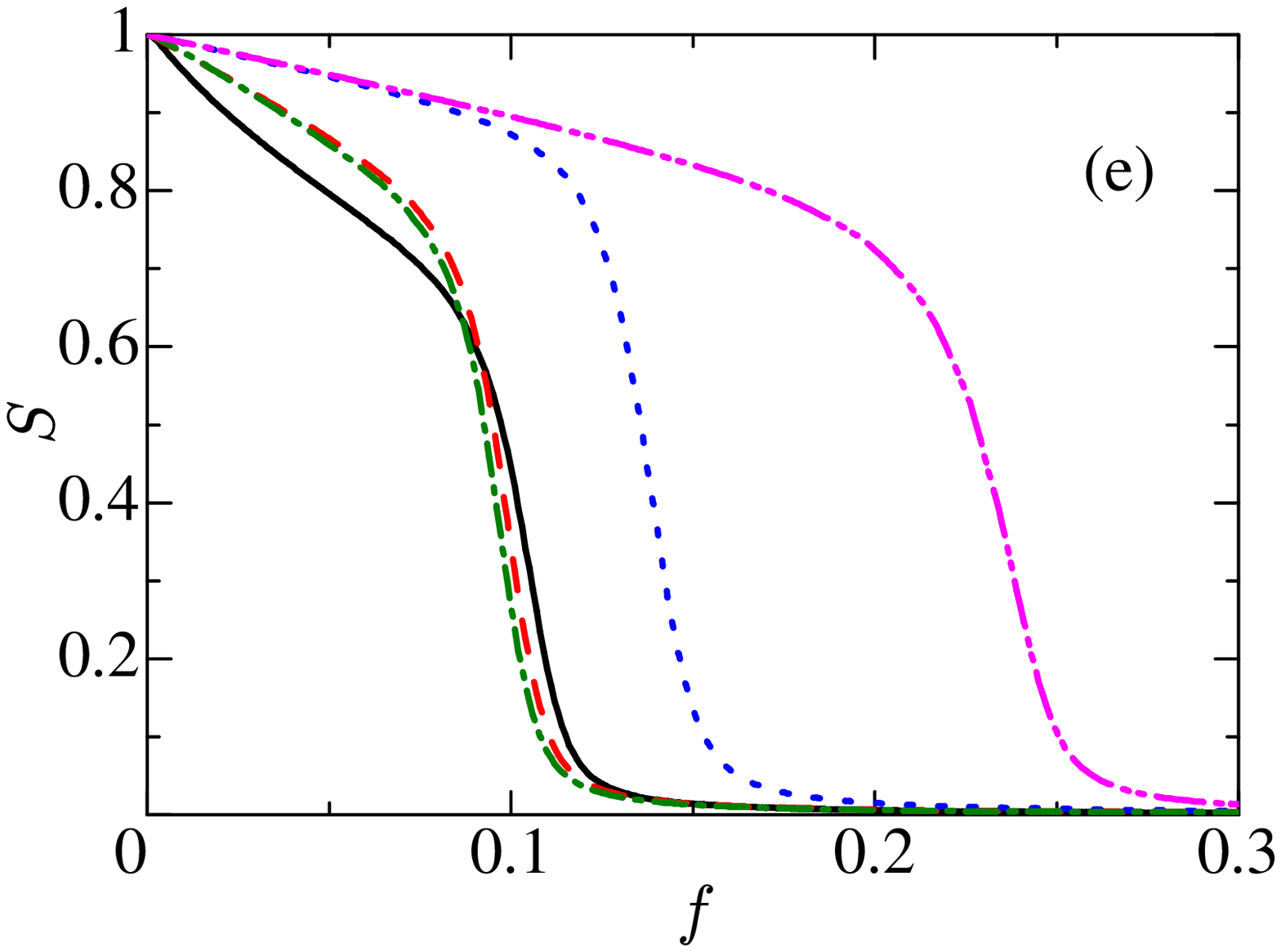}
		\end{tabular}
	\end{center}
	\caption{(Color online) The relative size $S$ of the largest connected component when a fraction $f$ of the nodes are removed from the network under the (a) RA (random attack), (b) SDA (static degree-based attack), (c) SBA (static betweenness-based attack), (d) DDA (dynamic degree-based attack), and (e) DBA (dynamic betweenness-based attack) strategies. Solid line represents the HL network; dashed line, the UN network; dotted line, the HH network; dash-dotted line, the PA network; and dash-double dotted line, the ER network. These lines are averaged over 1000 trials.}
	\label{fig:fig2}
\end{figure*}

\begin{figure*}[tbp]
	\begin{center}
		\begin{tabular}{c}
			\includegraphics[scale=0.45]{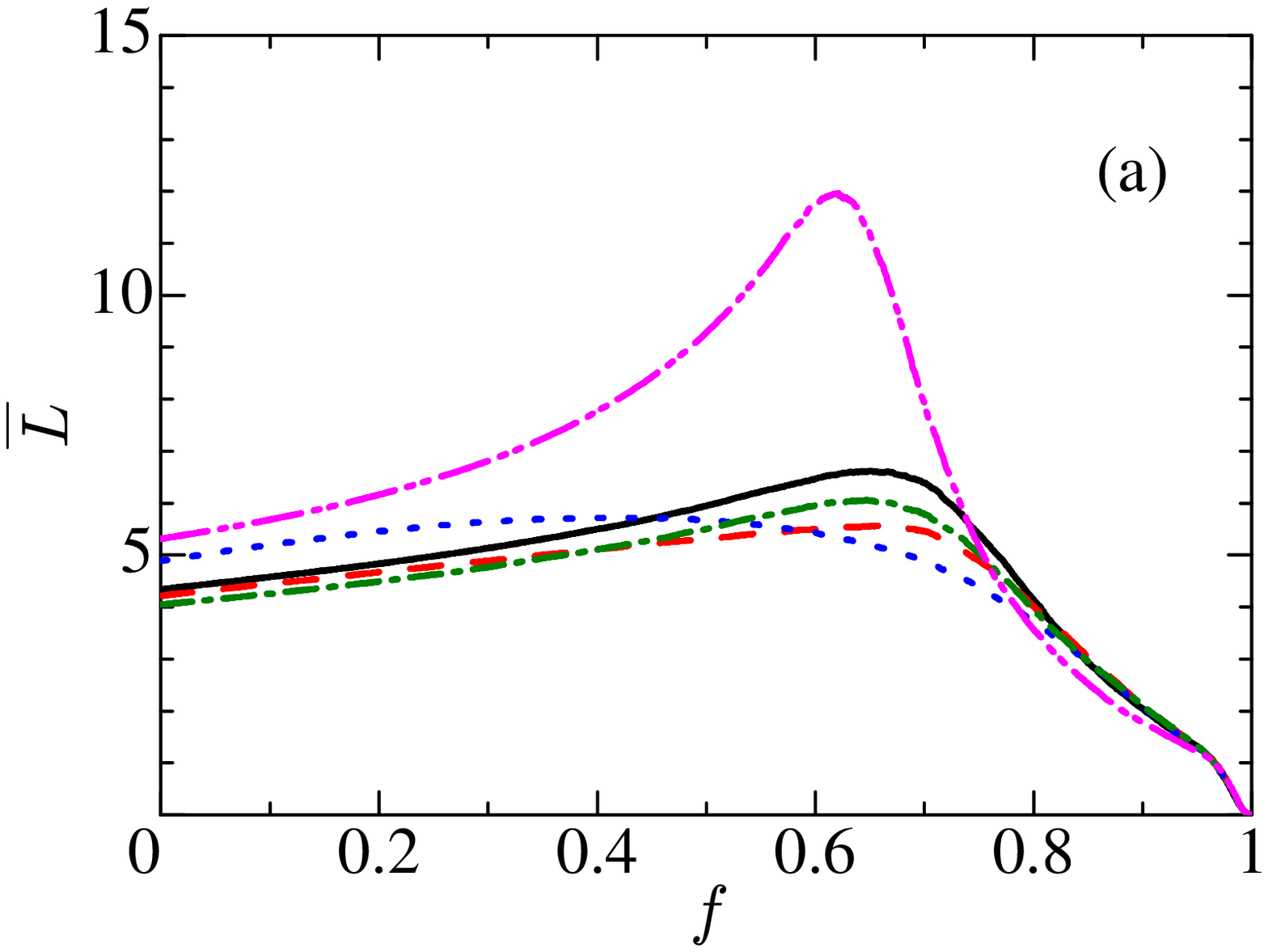}\ \ \ \ \ \ \ \ \
			\includegraphics[scale=0.45]{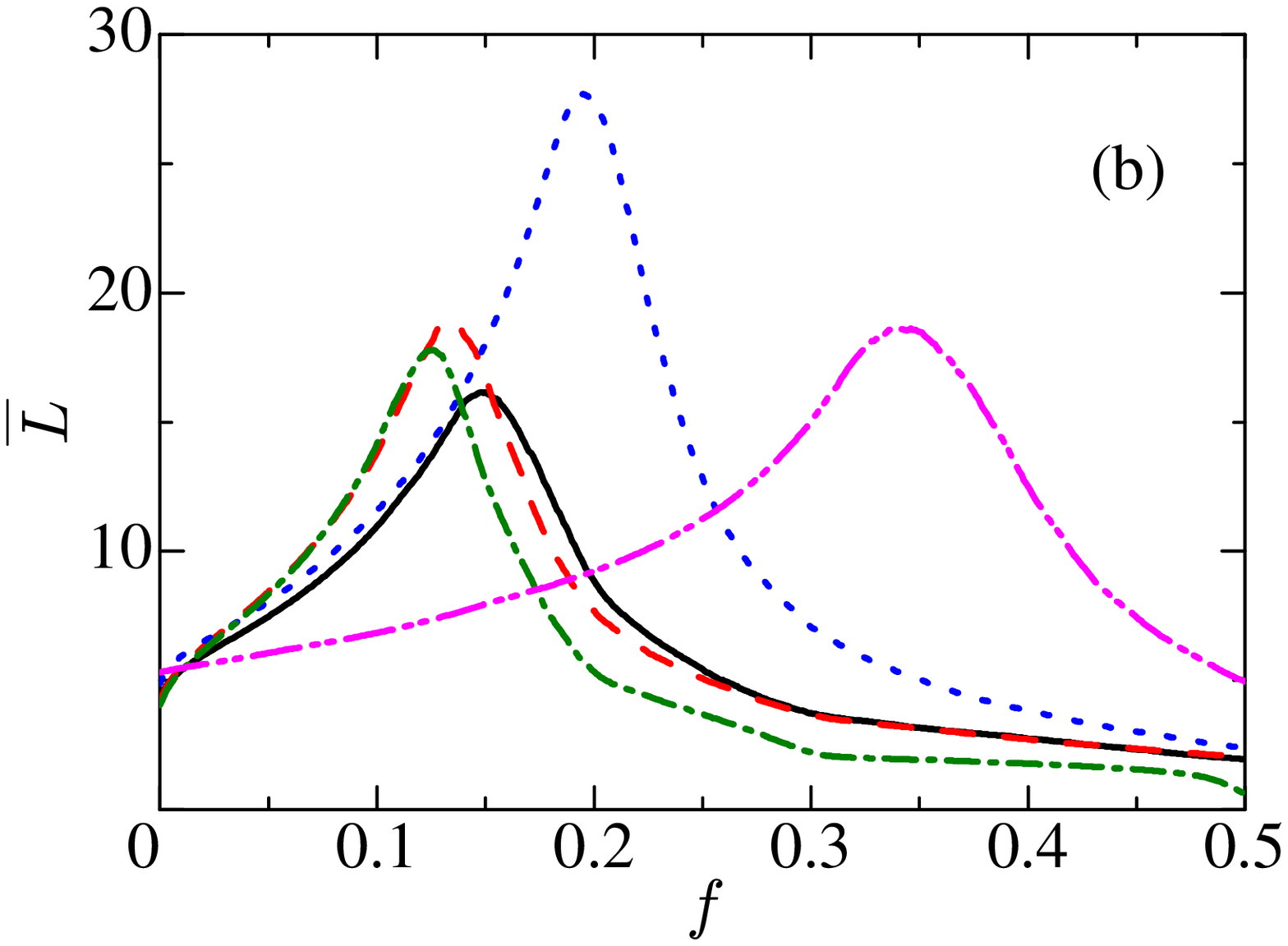}\\ \\
			\includegraphics[scale=0.45]{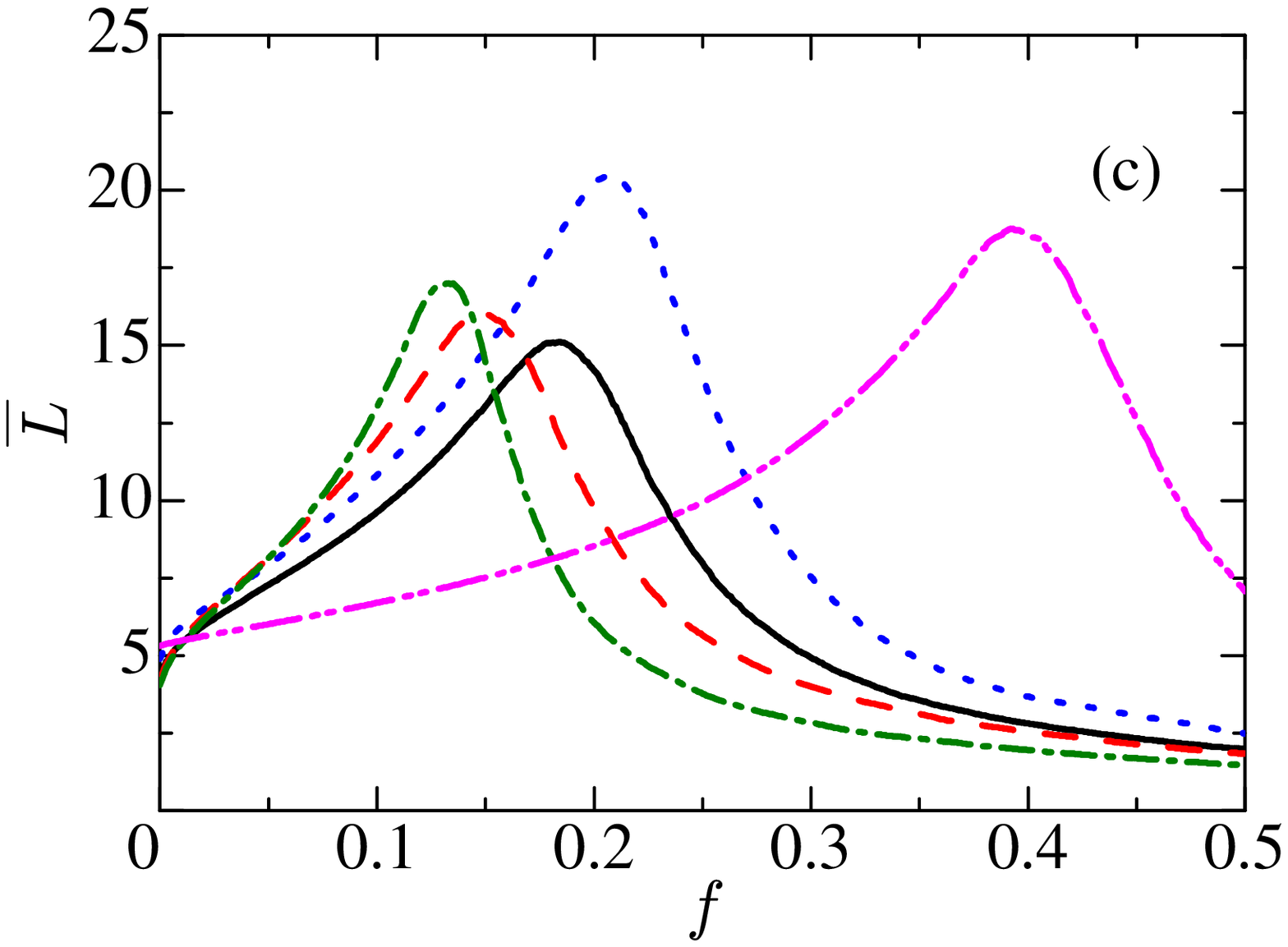}\ \ \ \ \ \ \ \ \
			\includegraphics[scale=0.45]{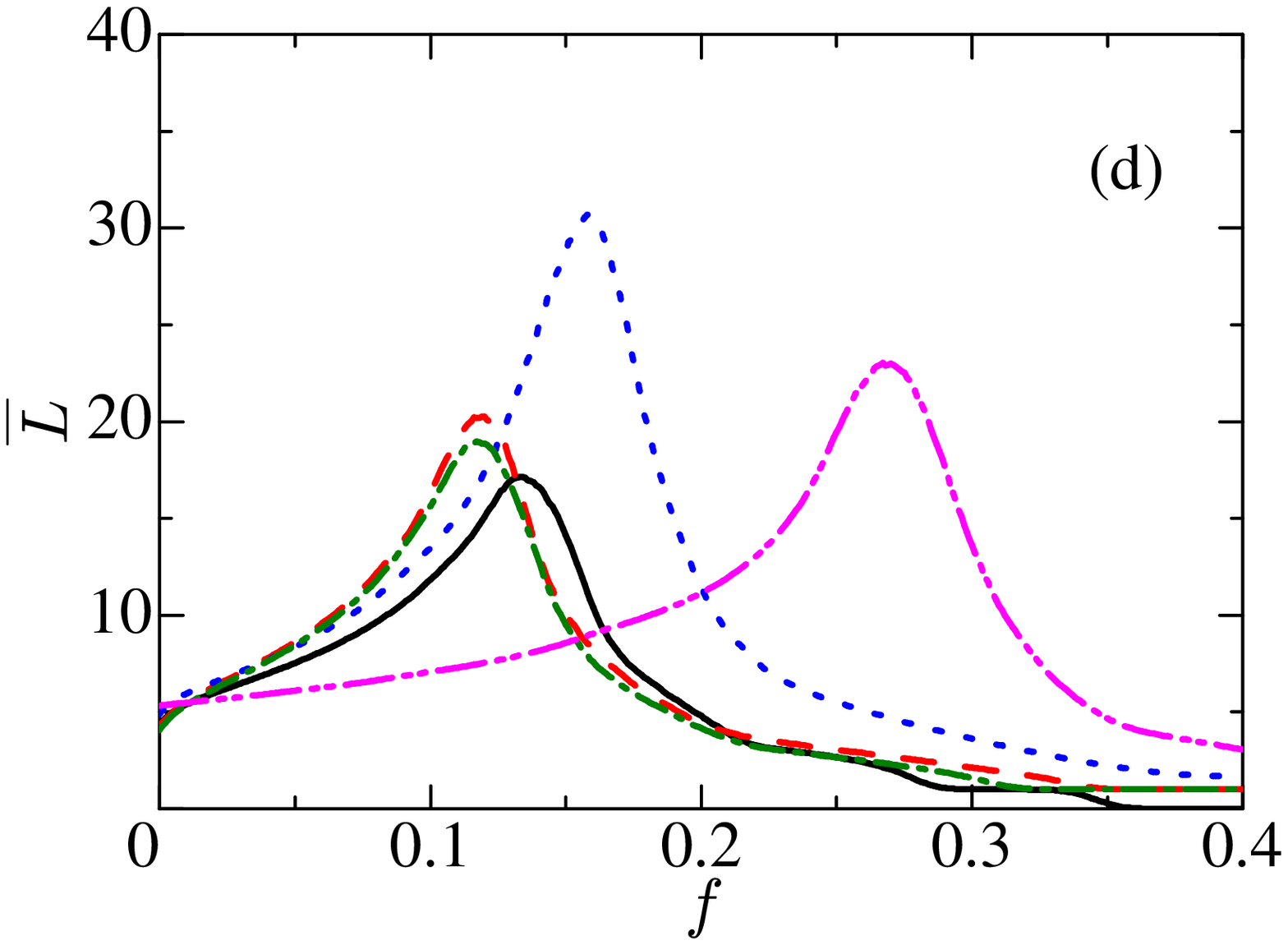}\\ \\
			\includegraphics[scale=0.45]{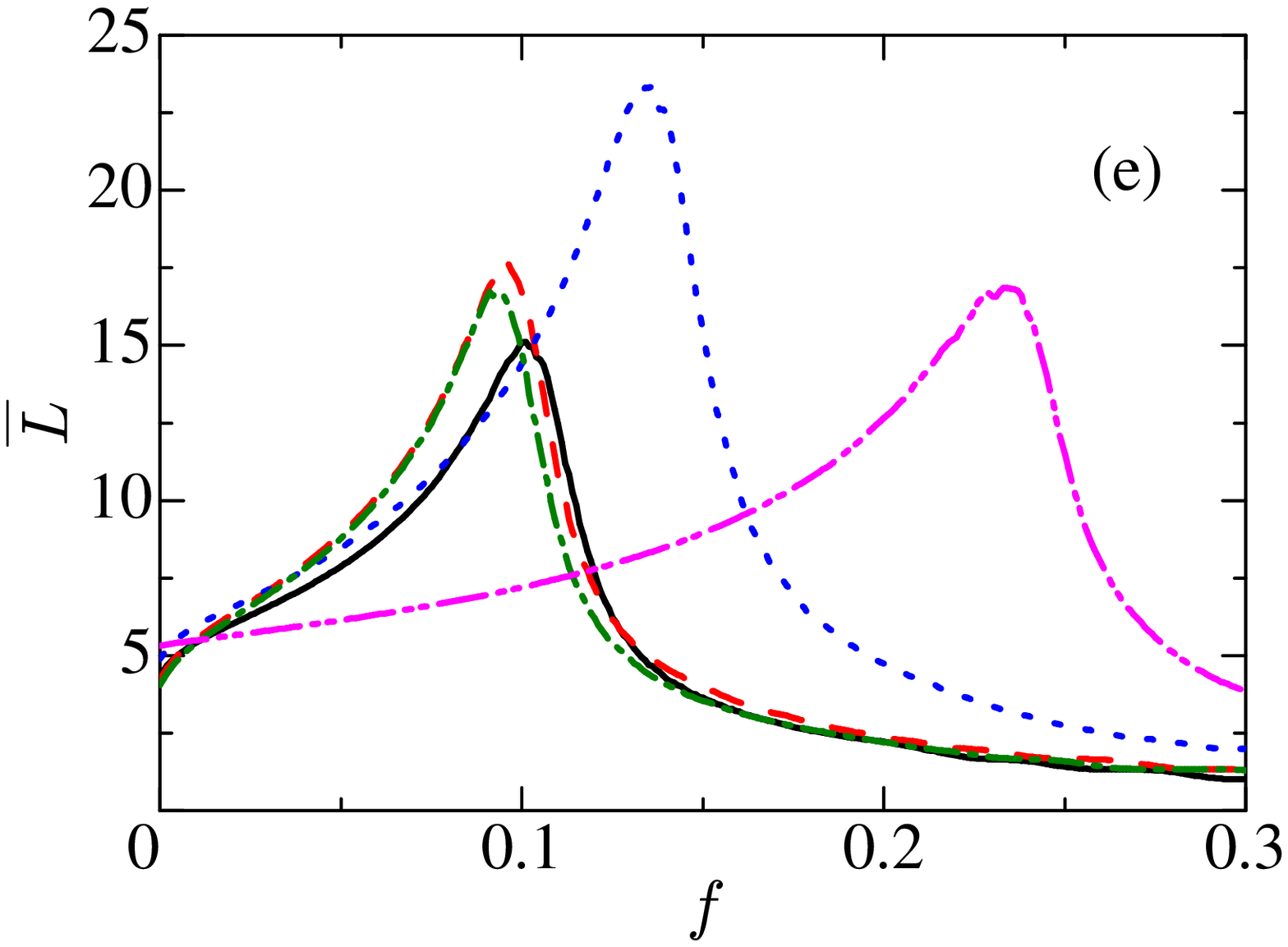}
		\end{tabular}
	\end{center}
	\caption{(Color online) The average path length $\overline{L}$ of the largest connected component when a fraction $f$ of the nodes are removed from the network under the (a) RA (random attack), (b) SDA (static degree-based attack), (c) SBA (static betweenness-based attack), (d) DDA (dynamic degree-based attack), and (e) DBA (dynamic betweenness-based attack) strategies. Solid line represents the HL network; dashed line, the UN network; dotted line, the HH network; dash-dotted line, the PA network; and dash-double dotted line, the ER network. These lines are averaged over 1000 trials.}
	\label{fig:fig3}
\end{figure*}

To determine the critical threshold $f_{c}$ at which a network is disrupted, we monitor the relative size $S$ and the average path length $\overline{L}$ of the largest connected component in function of the fraction $f$ of node removal, as shown in Fig \ref{fig:fig2} and Fig \ref{fig:fig3}.
The fraction of node removal represents $f=m/N$, where $m$ is the number of removed nodes and $N$ is the number of nodes in the initial connected network.
$S$ represents the fraction of nodes contained in the largest connected component out of the $N$ nodes.
As nodes are removed, $S$ decreases from $S=1$, and gradually the network breaks into isolated clusters.
At $S=0$ the network completely breaks apart.
Meanwhile, as nodes are removed, $\overline{L}$ increases and peaks just before the network is disrupted.
After the network breaks into isolated clusters, $\overline{L}$ decreases rapidly because the size of the largest connected component decreases drastically.
In this study, the critical threshold $f_{c}$ is determined by the fraction of removed nodes from the network when $\overline{L}$ is maximum, instead of using the point $S=0$, since small isolated clusters do not work as a system anymore.

Table \ref{tab:tab2} shows the critical threshold $f_{c}$ under all the attack strategies.
Large $f_c$ indicates that the network is robust in terms of connectivity.
For all the network models, the order of harmful attack strategy is RA $<$ SBA $<$ SDA $<$ DDA $<$ DBA (RA $<$ SBA means that SBA is more harmful than RA).
The order of robust network model is different whether the type of attack strategy is random or intentional.
Under the RA strategy, the order of robust network is HH $<$ ER $<$ UN $<$ PA $<$ HL (HH $<$ ER means that ER is more robust than HH), but under the four intentional strategies, it changes into PA $<$ UN $<$ HL $<$ HH $<$ ER.
From our numerical results, we find that the HL network we proposed is structurally more robust than the PA and UN networks under all the attack strategies.
It is surprising that the HH network is much more fragile than the ER network against random attacks although it is more robust than the HL network against intentional attacks.

Under the RA strategy, all the network models display the high robustness compared with the intentional attack strategies.
The critical threshold of the HL network is the highest, and that of the HH network is quite low against random attacks.
The network in which hubs are kept away from each other can further increase the robustness against random attacks, whereas dense interconnections between hubs make the network fragile.

Comparing the two static intentional attack strategies, the SDA is more harmful than the SBA for all the network models.
Since the betweenness centrality depends on the global network structure, the distribution of betweenness centrality varies substantially as nodes are removed from the network.
Thus, the initial value of betweenness centrality does not appropriately reflect the importance of node after several nodes are removed from the network.

The dynamic intentional attack strategies are more harmful than the static ones.
A large number of edges are removed faster under the DDA strategy, while the largest connected component is broken into isolated clusters faster under the DBA strategy.
As shown in Table \ref{tab:tab2}, the most lethal attack strategy is the DBA.
This result indicates that detecting the key nodes which connect many clusters is the most effective way for network disruption.

\subsection{Load tolerance}
\begin{figure*}[tbp]
	\begin{center}
		\begin{tabular}{c}
			\includegraphics[scale=0.45]{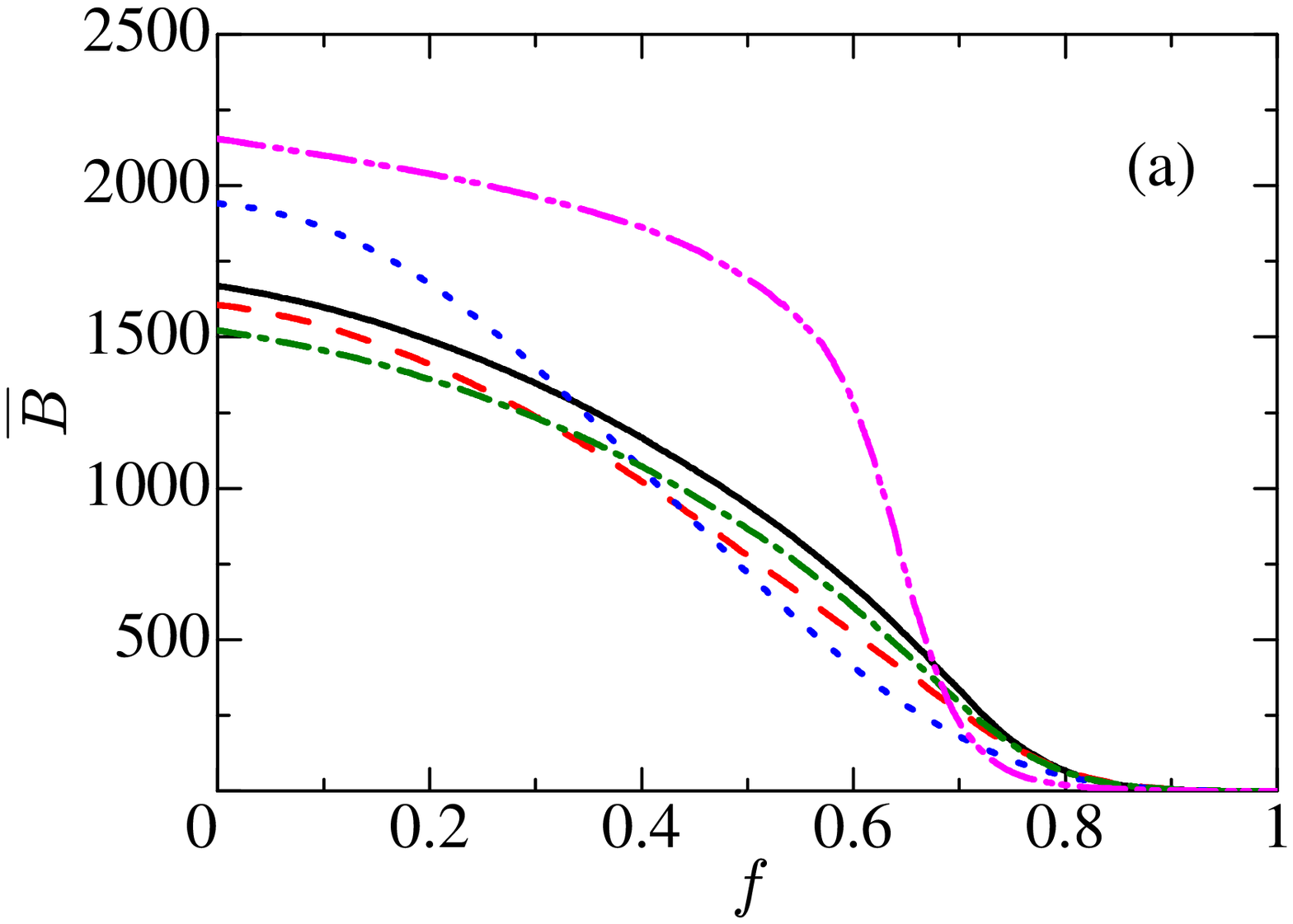}\ \ \ \ \ \ \ \ \
			\includegraphics[scale=0.45]{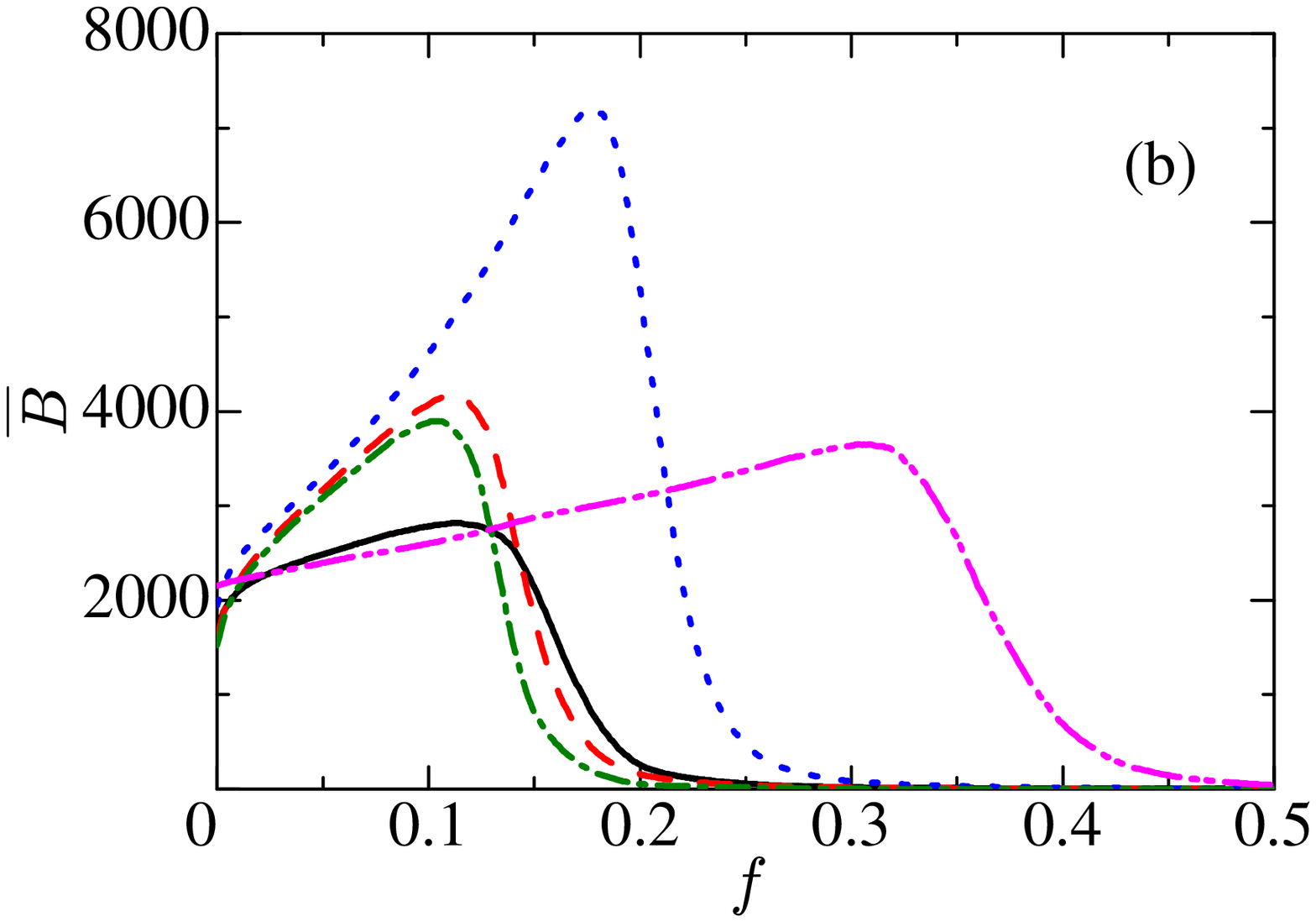}\\ \\
			\includegraphics[scale=0.45]{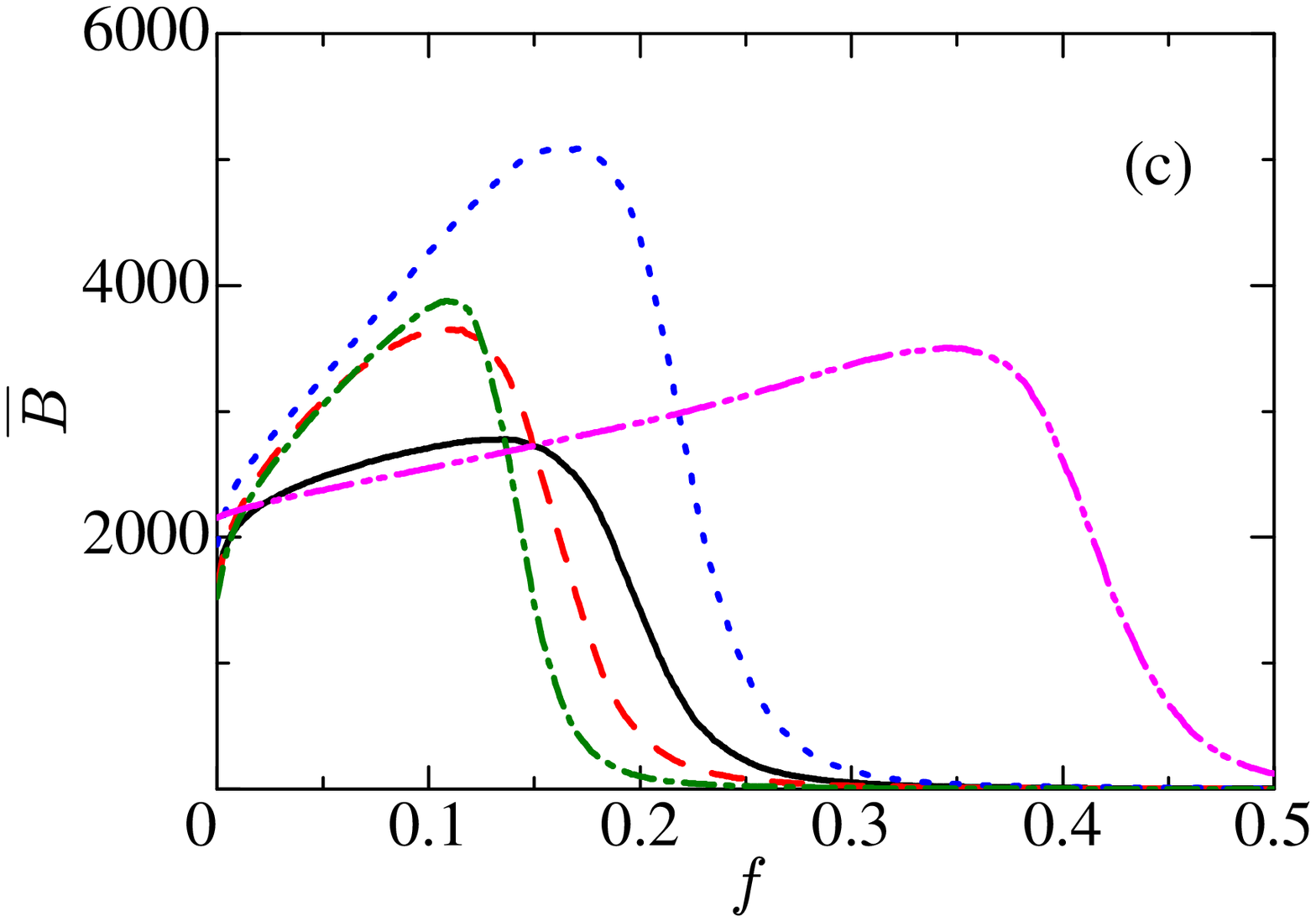}\ \ \ \ \ \ \ \ \
			\includegraphics[scale=0.45]{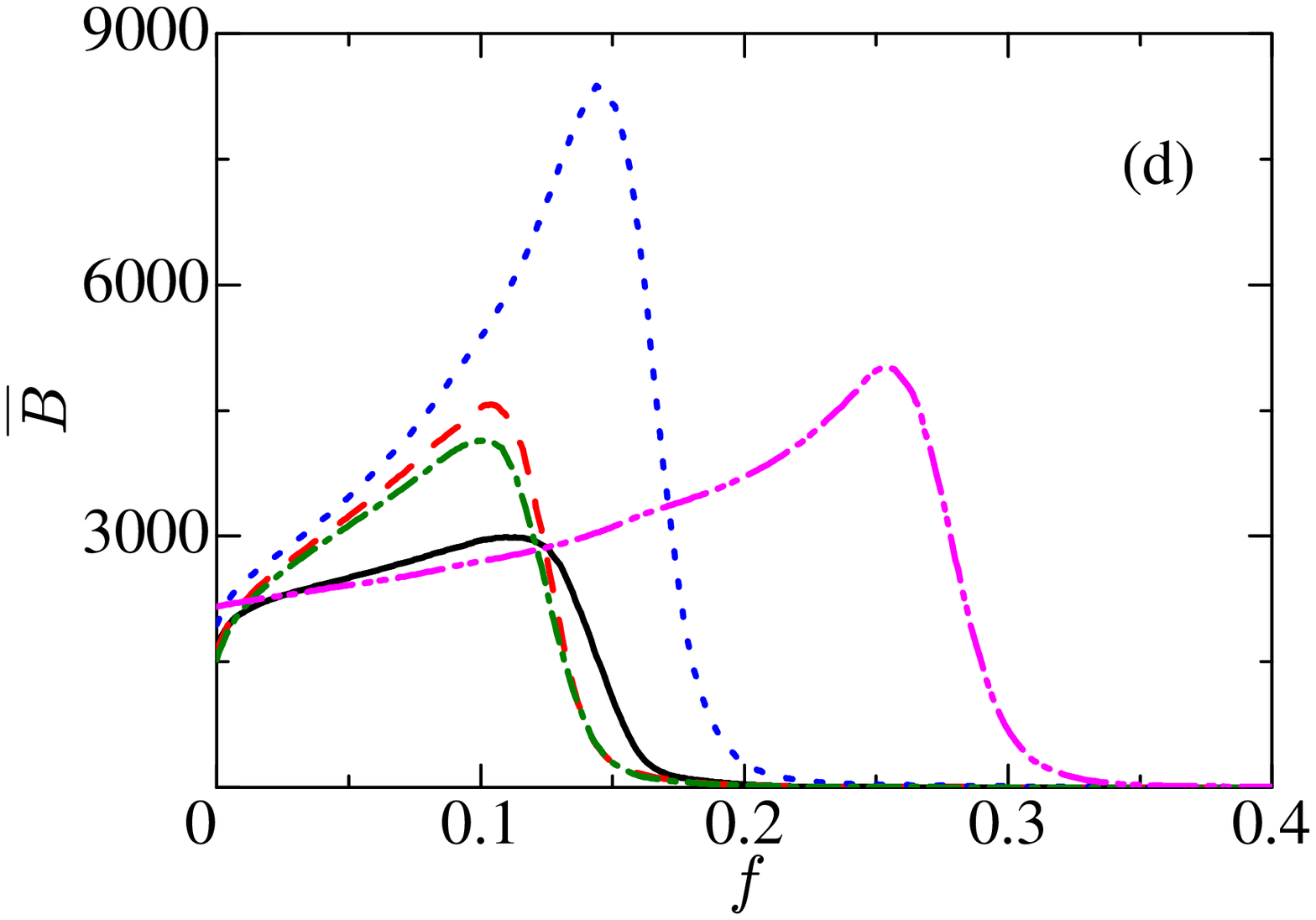}\\ \\
			\includegraphics[scale=0.45]{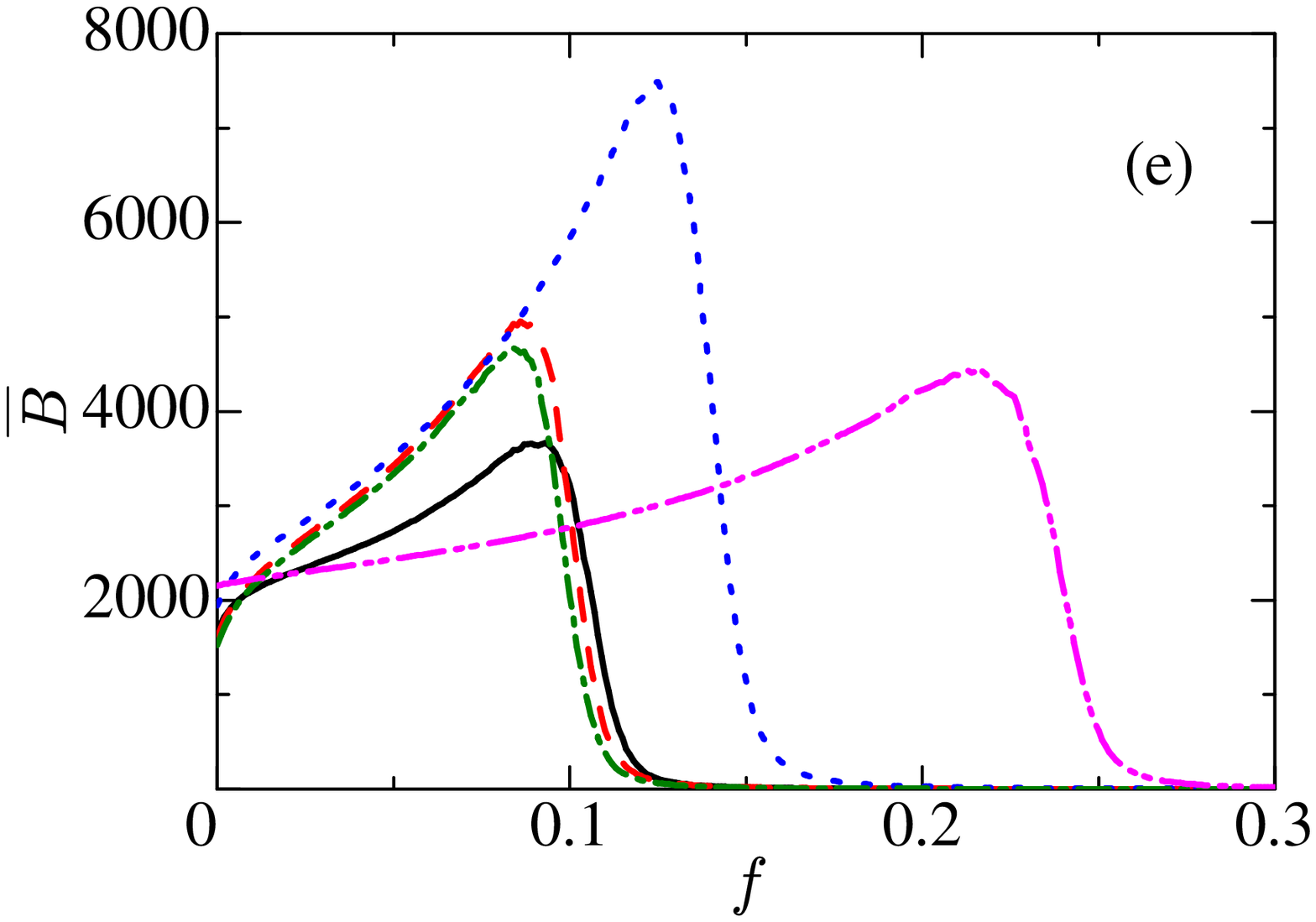}
		\end{tabular}
	\end{center}
	\caption{(Color online) The average betweenness centrality $\overline{B}$ of the largest connected component when a fraction $f$ of the nodes are removed from the network under the (a) RA (random attack), (b) SDA (static degree-based attack), (c) SBA (static betweenness-based attack), (d) DDA (dynamic degree-based attack), and (e) DBA (dynamic betweenness-based attack) strategies. Solid line represents the HL network; dashed line, the UN network; dotted line, the HH network; dash-dotted line, the PA network; and dash-double dotted line, the ER network. These lines are averaged over 1000 trials.}
	\label{fig:fig4}
\end{figure*}

\begin{figure*}[tbp]
	\begin{center}
		\begin{tabular}{c}
			\includegraphics[scale=0.45]{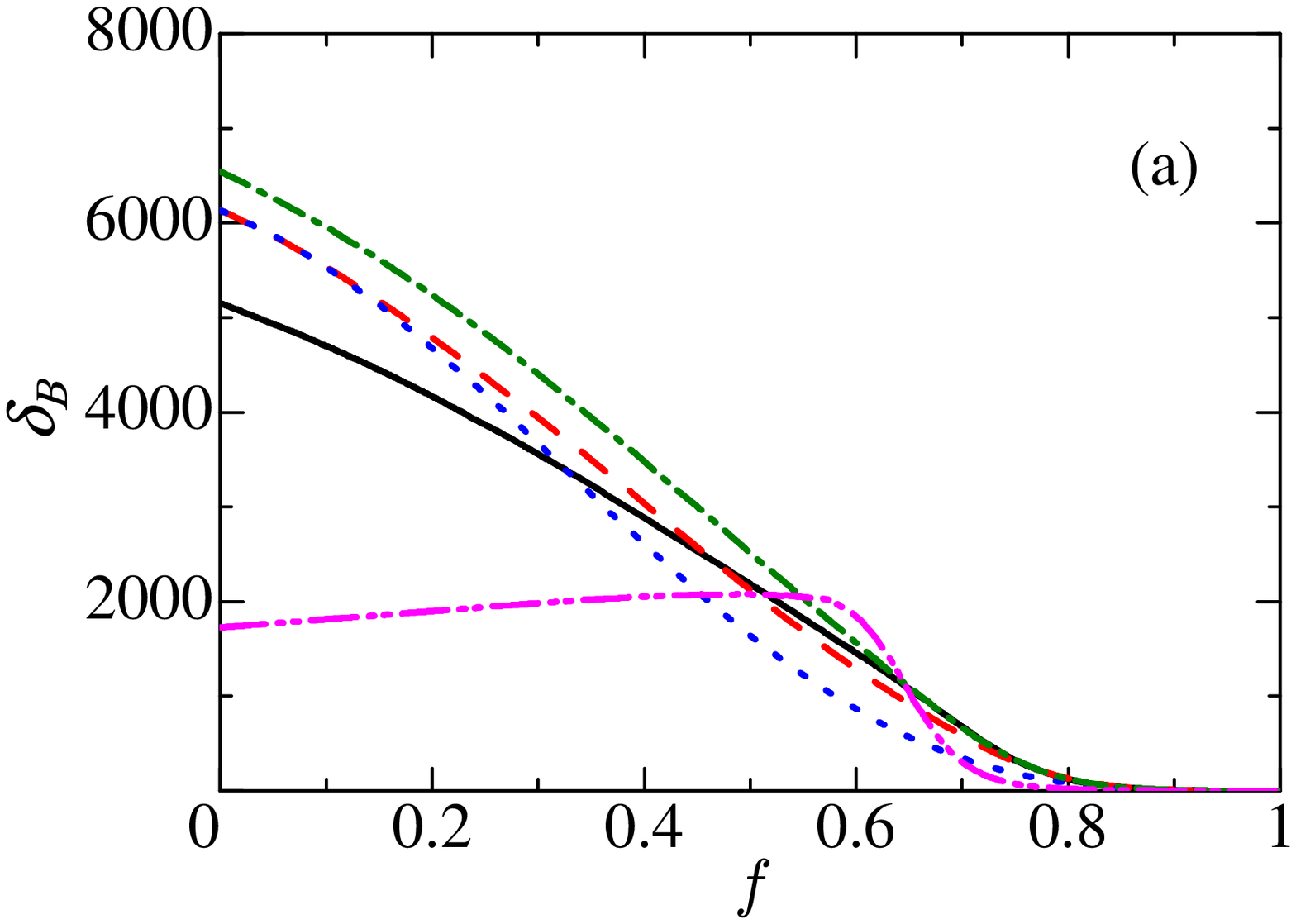}\ \ \ \ \ \ \ \ \
			\includegraphics[scale=0.45]{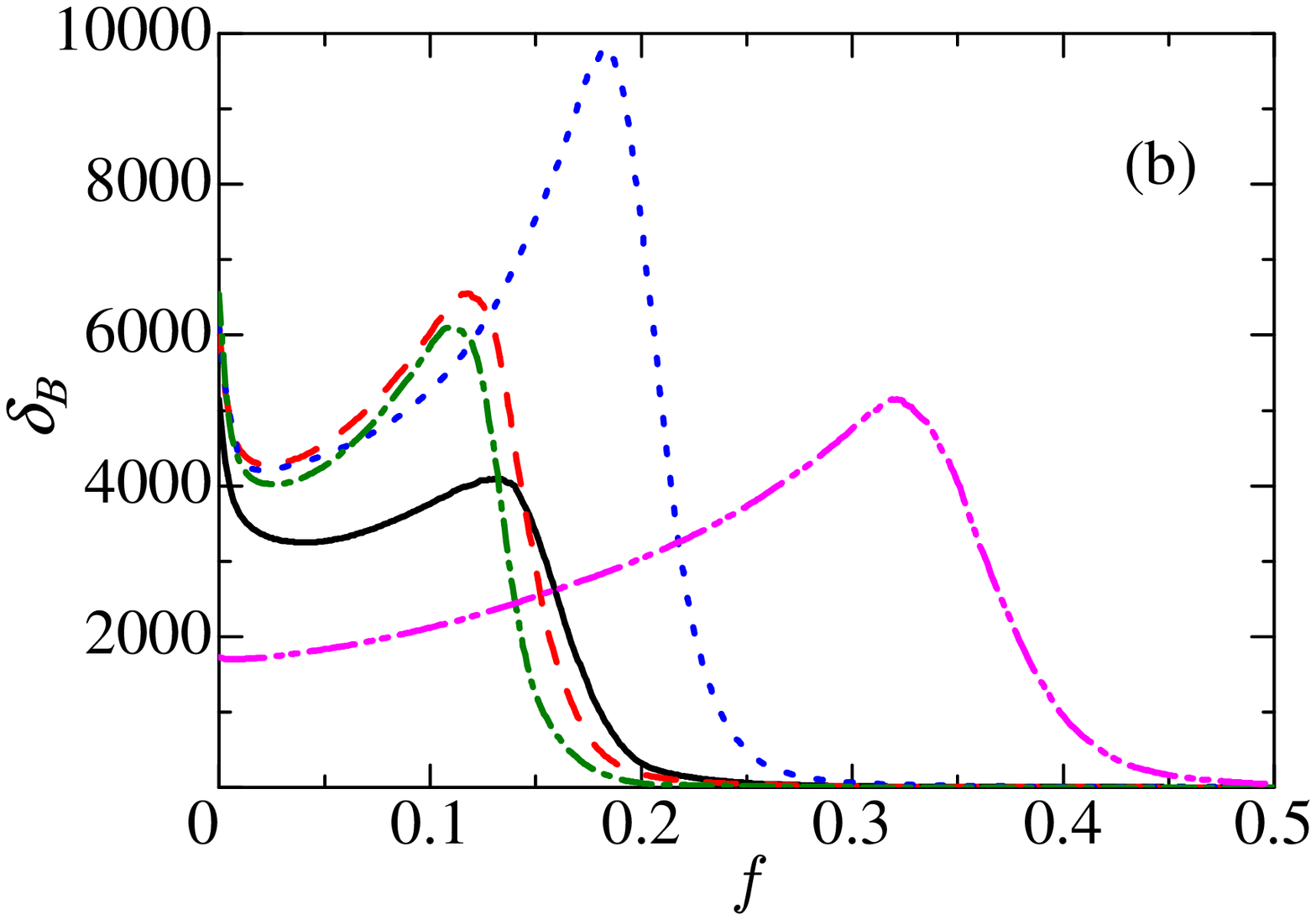}\\ \\
			\includegraphics[scale=0.45]{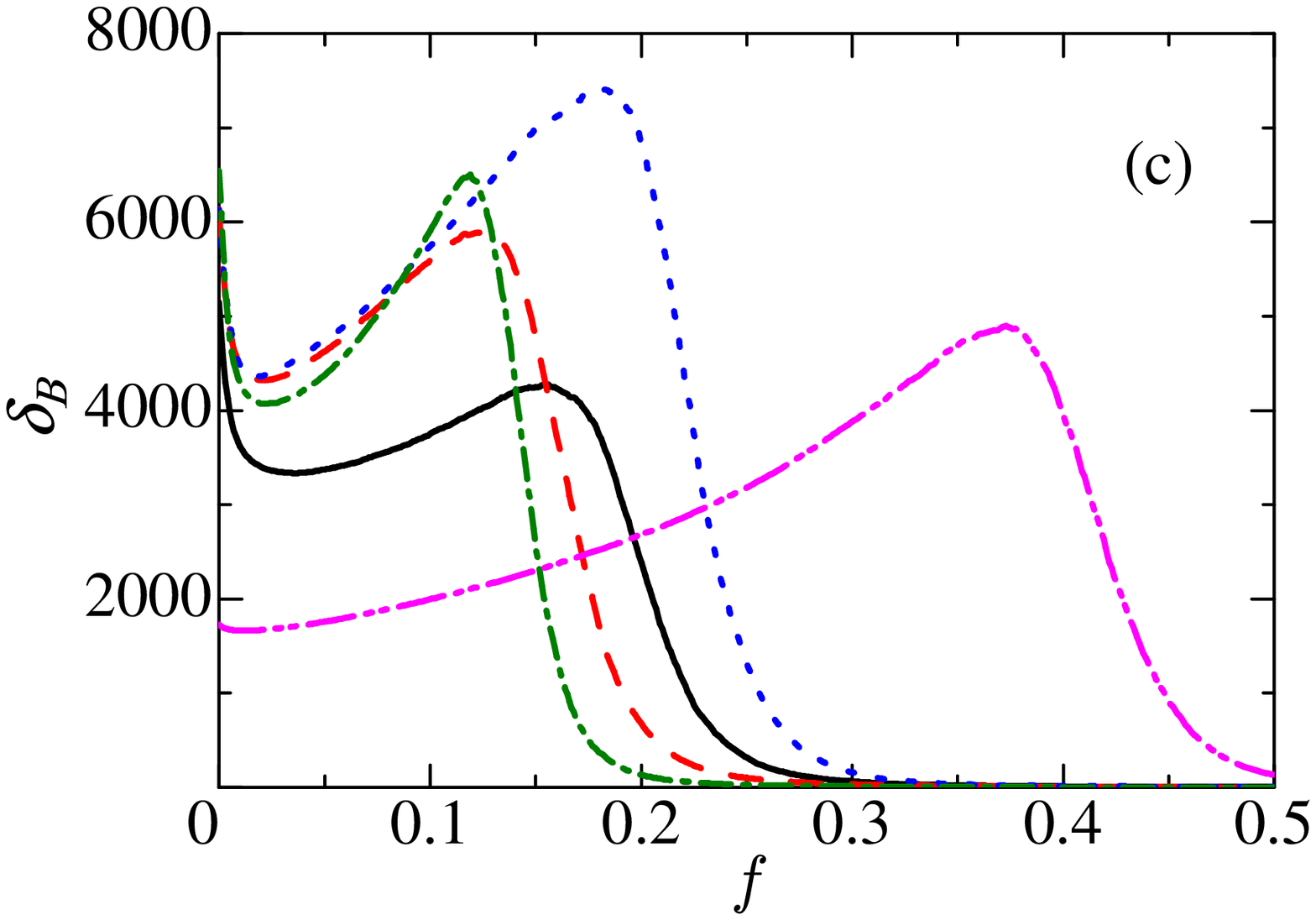}\ \ \ \ \ \ \ \ \
			\includegraphics[scale=0.45]{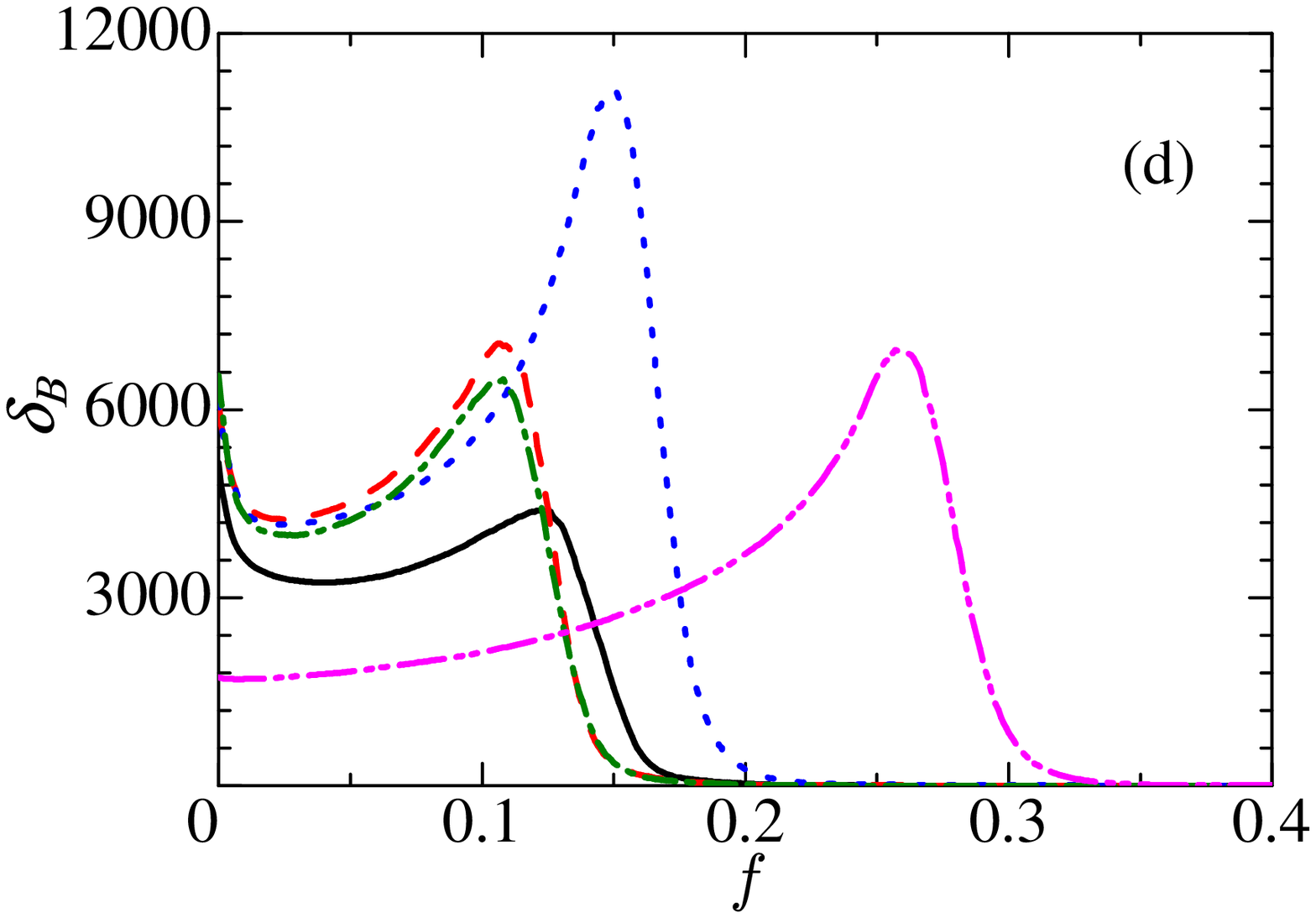}\\ \\
			\includegraphics[scale=0.45]{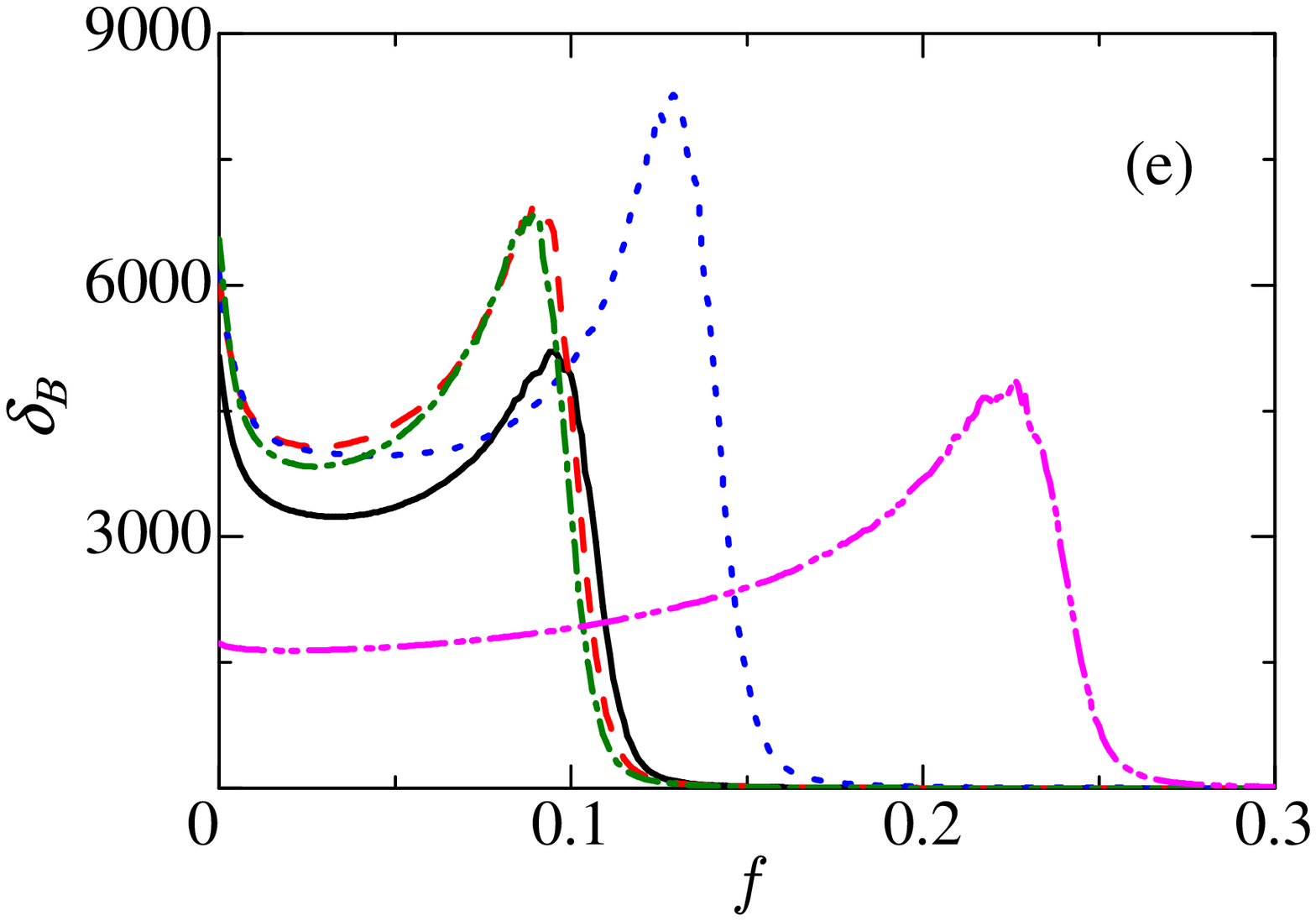}
		\end{tabular}
	\end{center}
	\caption{(Color online) The betweenness deviation $\delta_{B}$ of the largest connected component when a fraction $f$ of the nodes are removed from the network under the (a) RA (random attack), (b) SDA (static degree-based attack), (c) SBA (static betweenness-based attack), (d) DDA (dynamic degree-based attack), and (e) DBA (dynamic betweenness-based attack) strategies. Solid line represents the HL network; dashed line, the UN network; dotted line, the HH network; dash-dotted line, the PA network; and dash-double dotted line, the ER network. These lines are averaged over 1000 trials.}
	\label{fig:fig5}
\end{figure*}

To analyze the dynamic change of the amount of load and its distribution, we also monitor the average node betweenness centrality $\overline{B}$ and the betweenness deviation $\delta_{B}$ of the largest connected component, as shown in Fig \ref{fig:fig4} and Fig \ref{fig:fig5}.
Table \ref{tab:tab3} shows the fraction $f_{B}$ of node removal when the average betweenness centrality is maximum.
The maximum values $\overline{B}_{\max}$ and betweenness deviation $\delta_{B_{f_{B}}}$ at $f_{B}$ are shown in Table \ref{tab:tab4} and Table \ref{tab:tab5} respectively.

Under the RA strategy, $\overline{B}$ for all the networks decreases monotonically as nodes are removed.
$\delta_{B}$ for all the scale-free networks also decreases, while $\delta_{B}$ for the ER network keeps constant.
Thus $\overline{B}_{\max}$ and $\delta_{B_{f_{B}}}$ under the RA strategy are approximately the same as the values in the initial state.
This indicates that when random failures occur, the amount of load on networks and the degree of heterogeneity of load distribution do not increase.

Under the intentional attack strategies, $\overline{B}$ for the UN, HH, and PA networks increases rapidly as $f$ increases, and peaks at $f_{B}$ before the critical threshold $f_{c}$.
These results indicate that the breakdowns of nodes make the amount of load heavier when these networks are intentionally attacked.
After these networks break into isolated clusters, the average betweenness centrality $\overline{B}$ decreases rapidly as well as the average path length $\overline{L}$.
Betweenness deviation $\delta_{B}$ once drops by removal of a few nodes.
As nodes are removed, $\delta_{B}$ increases again, and peaks near $f_{B}$, indicating that the distribution of betweenness becomes heterogeneous by successive intentional attacks.
The increase in the heterogeneity of betweenness distribution indicates that the excessive load is concentrated on a few nodes; therefore, the UN, HH, and PA networks tend to cause the load congestion against intentional attacks.
For the HL network, both $\overline{B}$ and $\delta_{B}$ do not much increase when the nodes are removed successively except under the DBA strategy.
Even under the DBA strategy, the values of $\overline{B}$ and $\delta_{B}$ for the HL network are much smaller than those of the other scale-free networks.
Since the HL network can maintain the homogeneity of betweenness distribution against intentional attacks, it is a highly load-tolerant structure.
Although the maximum value of $\delta_{B}$ for the ER network is larger than that for the HL network, the ER network can maintain low value of $\delta_{B}$ longer.
This means that the distribution of betweenness centrality for the ER network is the most homogeneous.
The ER network is more load-tolerant than the scale-free networks.

\section{Discussions}
In this study, we have found that the topological assortativity strongly affects the robustness of scale-free networks.
Especially the HL network is more load-tolerant than the other models of scale-free networks.
The average degree of the networks used in this study is $\overline{k}=2E/N=4$; therefore, the results observed here with the HL model are applicable to the disassortative real networks such as the Internet ($\overline{k}=3.8$) \cite{vazquez2002} and the human protein interaction networks (4.624) \cite{rual2005}.
The common characteristic of these disassortative networks is that there exists dynamics such as traffic and interactions in the networks.
In communication networks, for example, the architectures which can avoid heavy concentration of traffic load are advantageous.
In biological networks, cells can robustly survive and retain their fundamental functional interactions under the perturbations by intrinsic errors and extrinsic stimulations.
Even if a large number of hubs are removed, the load intensity of the HL network does not increase excessively.
Our results indicate that the disassortative topology can avoid the propagation of lethal perturbations caused by removal of hubs.
This advantage of load tolerance for the HL network is not observed in the scale-free networks constructed with the PA model.

The HH network displays the connectivity robustness against intentional attacks; however, the amount of load and heterogeneity of the load distribution become extremely high.
Since excessive load intensity increases the possibility to cause the chain of node breakdowns, the HH network is not an appropriate structure for reliable systems.
The ER network also shows the connectivity robustness under the intentional strategies.
In addition, unlike the HH network, the degree of heterogeneity of the load distribution is quite low even if the important nodes are removed from the network.
Although the ER network has these advantages over the scale-free networks, many real networks adopt the scale-free structures.
The common disadvantage of both the HH and ER networks is the fragility of connectivity against random attacks and failures.
In real networks, random attacks are the most common perturbations compared with intentional attacks; therefore, these networks improve the connectivity robustness against random attacks with acquiring the disassortative topology.

In the non-assortative scale-free networks, the connectivity robustness of the UN network is similar to that of the PA network, but the load tolerance of the UN network is higher, as shown in Table \ref{tab:tab4} and Table \ref{tab:tab5}.
This indicates that the topology with the positive age correlation attenuates the load tolerance as in the case of the degree-degree correlation.

The HL model does not display high load tolerance only against the dynamic betweenness-based attack.
Because it is difficult to precisely search the node with the highest betweenness centrality without any calculations, the DBA strategy might be rarely observed in nature.
The order of the dynamic betweenness-based attacks corresponds with that of the sequential node breakdowns in the case where the network cannot withstand the heavy load.
The trigger of sequential node breakdowns seems to be an excessive increase of load when a network is attacked by other random or degree-based attack strategies.
By suppressing the excessive load concentration against these attacks, the disassortative topology avoids lethal shutdown of the whole system.

\section{Conclusion}
We have studied the connectivity robustness and load tolerance in the scale-free networks with different inner structures under the random and intentional attack strategies.
We have proposed the models of scale-free networks which can change the topological assortativity with fixing a degree distribution.
To conclude, we find that disassortative networks are structurally robust against random attacks and highly load-tolerant while assortative networks are most resistant to intentional attacks yet significantly fragile against random attacks.
The PA model has been widely used to study the basic functionalities of real networks; however, it is clear that this model is incomplete to account for the robustness of real networks.
The results presented here indicate other functionalities of scale-free networks may depend on the topological assortativity.
In the case of analyzing the properties of scale-free networks, their inner structures should be taken into consideration for the precise evaluation.

\begin{acknowledgments}
This work was supported by the Special Coordination Funds for Promoting Science and Technology from the Science and Technology Agency of Japan; the Strategic Information and Communications R\&D Promotion Programme of the Ministry of Internal Affairs and Communications; Japan Grants-in-Aid for Scientific Research 19319129 from Japan Society for the Promotion of Science; and Forerunner Pharma Research Co., Ltd.
\end{acknowledgments}

\begin{table*}[tbh]
\caption{Topological indices for the HL, UN, HH, PA, and ER networks (1000 nodes and 2000 edges) in the initial state: the degree exponent $\gamma$, assortativity $r$, average shortest path length $\overline{L}$, average node betweenness centrality $\overline{B}$, node betweenness deviation $\delta_{B}$, and average clustering coefficient $\overline{C}$. These values are averaged over 1000 trials, and their standard deviations are shown in parentheses.}
	\begin{ruledtabular}
		\begin{tabular}{ccccccc}
		& $\gamma$ & $r$ & $\overline{L}$ & $\overline{B}$ & $\delta_{B}$ & $\overline{C}$\\
		\hline
		HL & -2.309 (0.024) & -0.201 (0.025) & 4.342 (0.052) & 1669.184 (25.984) & 5150.485 (460.308) & 0.017 (0.005) \\
		UN & -2.309 (0.024) & -0.056 (0.016) & 4.216 (0.061) & 1606.285 (30.324) & 6136.614 (486.685) & 0.024 (0.005) \\
		HH & -2.309 (0.024) &  0.208 (0.074) & 4.887 (0.110) & 1941.549 (54.856) & 6133.316 (665.880) & 0.020 (0.003) \\
  		PA & -2.309 (0.024) & -0.083 (0.013) & 4.047 (0.058) & 1521.947 (29.105) & 6543.770 (502.933) & 0.029 (0.006) \\
		ER &  ---           & -0.008 (0.023) & 5.315 (0.015) & 2155.307 (7.642)  & 1725.676 (54.555) & 0.003 (0.001)
		\end{tabular}
	\end{ruledtabular}
	\label{tab:tab1}
\end{table*}
\begin{table*}[tbp]
\caption{The critical threshold $f_{c}$ for each network model under the RA (random attack), SDA (static degree-based attack), SBA (static betweenness-based attack), DDA (dynamic degree-based attack), and DBA (dynamic betweenness-based attack) strategies. These values are averaged over 1000 trials, and their standard deviations are shown in parentheses.}
	\begin{ruledtabular}
		\begin{tabular}{cccccc}
		&  \multicolumn{5}{c}{$f_{c}$}\\ \cline{2-6}
		& RA & SDA & SBA & DDA & DBA\\
		\hline
		HL & 0.670 (0.064) & 0.155 (0.016) & 0.190 (0.019) & 0.138 (0.011) & 0.103 (0.006)\\
  		UN & 0.642 (0.104) & 0.138 (0.013) & 0.156 (0.017) & 0.122 (0.009) & 0.098 (0.006)\\
		HH & 0.469 (0.141) & 0.200 (0.012) & 0.212 (0.018) & 0.161 (0.009) & 0.137 (0.007)\\
		PA & 0.655 (0.078) & 0.130 (0.011) & 0.136 (0.012) & 0.121 (0.009) & 0.096 (0.006)\\
		ER & 0.634 (0.033) & 0.352 (0.022) & 0.404 (0.022) & 0.273 (0.010) & 0.235 (0.008)
		\end{tabular}
	\end{ruledtabular}
	\label{tab:tab2}
\end{table*}
\begin{table*}[btp]
\caption{The fraction $f_{B}$ of node removal when the average node betweenness centrality peaks under the RA (random attack), SDA (static degree-based attack), SBA (static betweenness-based attack), DDA (dynamic degree-based attack), and DBA (dynamic betweenness-based attack) strategies. These values are averaged over 1000 trials, and their standard deviations are shown in parentheses.}
	\begin{ruledtabular}
		\begin{tabular}{cccccc}
		&  \multicolumn{5}{c}{$f_{B}$}\\ 
		\cline{2-6}
		& RA & SDA & SBA & DDA & DBA\\
		\hline
		HL & 0.019 (0.032) & 0.122 (0.022) & 0.145 (0.029) & 0.118 (0.015) & 0.096 (0.008)\\
  		UN & 0.015 (0.024) & 0.088 (0.020) & 0.081 (0.021) & 0.090 (0.015) & 0.088 (0.009)\\
		HH & 0.023 (0.031) & 0.183 (0.014) & 0.174 (0.022) & 0.150 (0.009) & 0.130 (0.008)\\
		PA & 0.026 (0.047) & 0.109 (0.014) & 0.116 (0.015) & 0.106 (0.011) & 0.090 (0.007)\\
		ER & 0.026 (0.102) & 0.313 (0.025) & 0.359 (0.032) & 0.258 (0.011) & 0.223 (0.012)
		\end{tabular}
	\end{ruledtabular}
	\label{tab:tab3}
\end{table*}
\begin{table*}[btp]
\caption{The maximum value of average node betweenness centrality $\overline{B}_{\max}$ of the largest connected component at $f_{B}$ for each network model under the RA (random attack), SDA (static degree-based attack), SBA (static betweenness-based attack), DDA (dynamic degree-based attack), and DBA (dynamic betweenness-based attack) strategies. These values are averaged over 1000 trials, and their standard deviations are shown in parentheses.}
	\begin{ruledtabular}
		\begin{tabular}{cccccc}
		&  \multicolumn{5}{c}{$\overline{B}_{\max}$}\\ 
		\cline{2-6}
		& RA & SDA & SBA & DDA & DBA\\
		\hline
		HL & 1683.334 (36.773)  & 3231.445 (466.164)  & 3155.943 (423.851) & 3498.506 (516.877) & 5431.222 (1135.440)\\
		UN & 1622.297 (42.248)  & 3328.901 (401.278)  & 3079.637 (264.438) & 3541.010 (455.957) & 5578.857 (1279.415)\\
		HH & 1961.184 (63.661)  & 8549.646 (1394.956) & 5862.128 (886.470) & 9970.531 (1714.246)& 11490.158 (2404.594)\\
		PA & 1541.016 (44.647)  & 4471.303 (648.140)  & 4478.449 (636.512) & 4848.303 (754.920) & 6775.231 (1356.674)\\
		ER & 2162.629 (37.696)  & 4181.165 (557.778)  & 4050.177 (565.841) & 5793.221 (869.380) & 6745.126 (1347.186)
		\end{tabular}
	\end{ruledtabular}
	\label{tab:tab4}
\end{table*}
\begin{table*}[btp]
\caption{The betweenness deviation $\delta_{B_{f_{B}}}$ of the largest connected component at $f_{B}$ shown in Table \ref{tab:tab3} for each network model under the RA (random attack), SDA (static degree-based attack), SBA (static betweenness-based attack), DDA (dynamic degree-based attack), and DBA (dynamic betweenness-based attack) strategies. These values are averaged over 1000 trials, and their standard deviations are shown in parentheses.}
	\begin{ruledtabular}
		\begin{tabular}{cccccc}
		&  \multicolumn{5}{c}{$\delta_{B_{f_{B}}}$}\\ 
		\cline{2-6}
		& RA & SDA & SBA & DDA & DBA\\
		\hline
		HL & 5051.387 (498.952) & 5031.664 (1229.997)  & 5192.662 (1248.097) & 5504.870 (1315.247)  & 10652.721 (3195.809)\\
		UN & 6051.705 (529.403) & 6002.702 (1580.187)  & 5385.555 (1050.862) & 6541.205 (1706.036)  & 12234.777 (4166.939)\\
		HH & 6014.938 (692.899) & 12621.442 (3195.688) & 8914.047 (2109.814) & 14609.197 (3759.824) & 19659.712 (5828.889)\\
		PA & 6333.455 (669.259) & 7493.152 (1812.646)  & 7987.298 (1875.853) & 8197.516 (2038.289)  & 13485.849 (3838.336)\\
		ER & 1798.551 (350.141) & 6231.814 (1531.553)  & 6029.592 (1545.539) & 8676.582 (2184.394)  & 12136.679 (3814.250)
		\end{tabular}
	\end{ruledtabular}
	\label{tab:tab5}
\end{table*}



\begin{thebibliography}{30}
\expandafter\ifx\csname natexlab\endcsname\relax\def\natexlab#1{#1}\fi
\expandafter\ifx\csname bibnamefont\endcsname\relax
  \def\bibnamefont#1{#1}\fi
\expandafter\ifx\csname bibfnamefont\endcsname\relax
  \def\bibfnamefont#1{#1}\fi
\expandafter\ifx\csname citenamefont\endcsname\relax
  \def\citenamefont#1{#1}\fi
\expandafter\ifx\csname url\endcsname\relax
  \def\url#1{\texttt{#1}}\fi
\expandafter\ifx\csname urlprefix\endcsname\relax\def\urlprefix{URL }\fi
\providecommand{\bibinfo}[2]{#2}
\providecommand{\eprint}[2][]{\url{#2}}

\bibitem[{\citenamefont{Albert and Barab\'asi}(2002)}]{albert2002}
\bibinfo{author}{\bibfnamefont{R.}~\bibnamefont{Albert}} \bibnamefont{and}
  \bibinfo{author}{\bibfnamefont{A.-L.} \bibnamefont{Barab\'asi}},
  \bibinfo{journal}{Rev. Mod. Phys.} \textbf{\bibinfo{volume}{74}},
  \bibinfo{pages}{47} (\bibinfo{year}{2002}).

\bibitem[{\citenamefont{Dorogovtsev and Mendes}(2003)}]{dorogovtsev2003}
\bibinfo{author}{\bibfnamefont{S.~N.} \bibnamefont{Dorogovtsev}}
  \bibnamefont{and} \bibinfo{author}{\bibfnamefont{J.~F.~F.}
  \bibnamefont{Mendes}}, \emph{\bibinfo{title}{Evolution of Networks: From
  Biology to the Internet and the WWW}} (\bibinfo{publisher}{Oxford University
  Press}, \bibinfo{address}{Oxford}, \bibinfo{year}{2003}).

\bibitem[{\citenamefont{Callaway et~al.}(2000)\citenamefont{Callaway, Newman,
  Strogatz, and Watts}}]{callaway2000}
\bibinfo{author}{\bibfnamefont{D.~S.} \bibnamefont{Callaway}},
  \bibinfo{author}{\bibfnamefont{M.~E.~J.} \bibnamefont{Newman}},
  \bibinfo{author}{\bibfnamefont{S.~H.} \bibnamefont{Strogatz}},
  \bibnamefont{and} \bibinfo{author}{\bibfnamefont{D.~J.} \bibnamefont{Watts}},
  \bibinfo{journal}{Phys. Rev. Lett} \textbf{\bibinfo{volume}{85}},
  \bibinfo{pages}{5468} (\bibinfo{year}{2000}).

\bibitem[{\citenamefont{Cohen et~al.}(2000)\citenamefont{Cohen, Erez,
  ben-Avraham, and Havlin}}]{cohen2000}
\bibinfo{author}{\bibfnamefont{R.}~\bibnamefont{Cohen}},
  \bibinfo{author}{\bibfnamefont{K.}~\bibnamefont{Erez}},
  \bibinfo{author}{\bibfnamefont{D.}~\bibnamefont{ben-Avraham}},
  \bibnamefont{and} \bibinfo{author}{\bibfnamefont{S.}~\bibnamefont{Havlin}},
  \bibinfo{journal}{Phys. Rev. Lett} \textbf{\bibinfo{volume}{85}},
  \bibinfo{pages}{4626} (\bibinfo{year}{2000}).

\bibitem[{\citenamefont{Cohen et~al.}(2001)\citenamefont{Cohen, Erez,
  ben-Avraham, and Havlin}}]{cohen2001}
\bibinfo{author}{\bibfnamefont{R.}~\bibnamefont{Cohen}},
  \bibinfo{author}{\bibfnamefont{K.}~\bibnamefont{Erez}},
  \bibinfo{author}{\bibfnamefont{D.}~\bibnamefont{ben-Avraham}},
  \bibnamefont{and} \bibinfo{author}{\bibfnamefont{S.}~\bibnamefont{Havlin}},
  \bibinfo{journal}{Phys. Rev. Lett} \textbf{\bibinfo{volume}{86}},
  \bibinfo{pages}{3682} (\bibinfo{year}{2001}).

\bibitem[{\citenamefont{Xulvi-Brunet et~al.}(2003)\citenamefont{Xulvi-Brunet,
  Pietsch, and Sokolov}}]{xulvi-brunet2003}
\bibinfo{author}{\bibfnamefont{R.}~\bibnamefont{Xulvi-Brunet}},
  \bibinfo{author}{\bibfnamefont{W.}~\bibnamefont{Pietsch}}, \bibnamefont{and}
  \bibinfo{author}{\bibfnamefont{I.~M.} \bibnamefont{Sokolov}},
  \bibinfo{journal}{Phys. Rev. E} \textbf{\bibinfo{volume}{68}},
  \bibinfo{pages}{036119} (\bibinfo{year}{2003}).

\bibitem[{\citenamefont{Serrano and Bogu\~n\'a}(2006{\natexlab{a}})}]{serrano2006-l}
\bibinfo{author}{\bibfnamefont{M.~A.} \bibnamefont{Serrano}} \bibnamefont{and}
  \bibinfo{author}{\bibfnamefont{M.}~\bibnamefont{Bogu\~n\'a}},
  \bibinfo{journal}{Phys. Rev. Lett} \textbf{\bibinfo{volume}{97}},
  \bibinfo{pages}{088701} (\bibinfo{year}{2006}{\natexlab{a}}).

\bibitem[{\citenamefont{Serrano and Bogu\~n\'a}(2006{\natexlab{b}})}]{serrano2006-2}
\bibinfo{author}{\bibfnamefont{M.~A.} \bibnamefont{Serrano}} \bibnamefont{and}
  \bibinfo{author}{\bibfnamefont{M.}~\bibnamefont{Bogu\~n\'a}},
  \bibinfo{journal}{Phys. Rev. E} \textbf{\bibinfo{volume}{74}},
  \bibinfo{pages}{056115} (\bibinfo{year}{2006}{\natexlab{b}}).

\bibitem[{\citenamefont{Albert et~al.}(2000)\citenamefont{Albert, Jeong, and
  Barab\'asi}}]{albert2000}
\bibinfo{author}{\bibfnamefont{R.}~\bibnamefont{Albert}},
  \bibinfo{author}{\bibfnamefont{H.}~\bibnamefont{Jeong}}, \bibnamefont{and}
  \bibinfo{author}{\bibfnamefont{A.-L.} \bibnamefont{Barab\'asi}},
  \bibinfo{journal}{Nature (London)} \textbf{\bibinfo{volume}{406}},
  \bibinfo{pages}{378} (\bibinfo{year}{2000}).

\bibitem[{\citenamefont{Gallos et~al.}(2005)\citenamefont{Gallos, Cohen,
  Argyrakis, Bunde, and Havlin}}]{gallos2005}
\bibinfo{author}{\bibfnamefont{L.~K.} \bibnamefont{Gallos}},
  \bibinfo{author}{\bibfnamefont{R.}~\bibnamefont{Cohen}},
  \bibinfo{author}{\bibfnamefont{P.}~\bibnamefont{Argyrakis}},
  \bibinfo{author}{\bibfnamefont{A.}~\bibnamefont{Bunde}}, \bibnamefont{and}
  \bibinfo{author}{\bibfnamefont{S.}~\bibnamefont{Havlin}},
  \bibinfo{journal}{Phys. Rev. Lett} \textbf{\bibinfo{volume}{94}},
  \bibinfo{pages}{188701} (\bibinfo{year}{2005}).

\bibitem[{\citenamefont{Holme et~al.}(2002)\citenamefont{Holme, Kim, Yoon, and
  Han}}]{holme2002}
\bibinfo{author}{\bibfnamefont{P.}~\bibnamefont{Holme}},
  \bibinfo{author}{\bibfnamefont{B.~J.} \bibnamefont{Kim}},
  \bibinfo{author}{\bibfnamefont{C.~N.} \bibnamefont{Yoon}}, \bibnamefont{and}
  \bibinfo{author}{\bibfnamefont{S.~K.} \bibnamefont{Han}},
  \bibinfo{journal}{Phys. Rev. E} \textbf{\bibinfo{volume}{65}},
  \bibinfo{pages}{056109} (\bibinfo{year}{2002}).

\bibitem[{\citenamefont{Barab\'asi and Albert}(1999)}]{barabasi1999}
\bibinfo{author}{\bibfnamefont{A.-L.} \bibnamefont{Barab\'asi}}
  \bibnamefont{and} \bibinfo{author}{\bibfnamefont{R.}~\bibnamefont{Albert}},
  \bibinfo{journal}{Science} \textbf{\bibinfo{volume}{286}},
  \bibinfo{pages}{509} (\bibinfo{year}{1999}).

\bibitem[{\citenamefont{Newman}(2002)}]{newman2002}
\bibinfo{author}{\bibfnamefont{M.~E.~J.} \bibnamefont{Newman}},
  \bibinfo{journal}{Phys. Rev. Lett} \textbf{\bibinfo{volume}{89}},
  \bibinfo{pages}{208701} (\bibinfo{year}{2002}).

\bibitem[{\citenamefont{Oshida and Ihara}(2006)}]{oshida2006}
\bibinfo{author}{\bibfnamefont{N.}~\bibnamefont{Oshida}} \bibnamefont{and}
  \bibinfo{author}{\bibfnamefont{S.}~\bibnamefont{Ihara}},
  \bibinfo{journal}{Phys. Rev. E} \textbf{\bibinfo{volume}{74}},
  \bibinfo{pages}{026115} (\bibinfo{year}{2006}).

\bibitem[{\citenamefont{Moreno et~al.}(2002)\citenamefont{Moreno, G\'omez, and
  Pacheco}}]{moreno2002}
\bibinfo{author}{\bibfnamefont{Y.}~\bibnamefont{Moreno}},
  \bibinfo{author}{\bibfnamefont{J.~B.} \bibnamefont{G\'omez}},
  \bibnamefont{and} \bibinfo{author}{\bibfnamefont{A.~F.}
  \bibnamefont{Pacheco}}, \bibinfo{journal}{Europhys. Lett.}
  \textbf{\bibinfo{volume}{58}}, \bibinfo{pages}{630} (\bibinfo{year}{2002}).

\bibitem[{\citenamefont{Motter and Lai}(2002)}]{motter2002}
\bibinfo{author}{\bibfnamefont{A.~E.} \bibnamefont{Motter}} \bibnamefont{and}
  \bibinfo{author}{\bibfnamefont{Y.~C.} \bibnamefont{Lai}},
  \bibinfo{journal}{Phys. Rev. E} \textbf{\bibinfo{volume}{66}},
  \bibinfo{pages}{065102(R)} (\bibinfo{year}{2002}).

\bibitem[{\citenamefont{Zhao et~al.}(2005)\citenamefont{Zhao, Lai, Park, and
  Ye}}]{zhao2005}
\bibinfo{author}{\bibfnamefont{L.}~\bibnamefont{Zhao}},
  \bibinfo{author}{\bibfnamefont{Y.-C.} \bibnamefont{Lai}},
  \bibinfo{author}{\bibfnamefont{K.}~\bibnamefont{Park}}, \bibnamefont{and}
  \bibinfo{author}{\bibfnamefont{N.}~\bibnamefont{Ye}}, \bibinfo{journal}{Phys.
  Rev. E} \textbf{\bibinfo{volume}{71}}, \bibinfo{pages}{026125}
  (\bibinfo{year}{2005}).

\bibitem[{\citenamefont{Lee et~al.}(2005)\citenamefont{Lee, Goh, Kahng, and
  Kim}}]{lee2005}
\bibinfo{author}{\bibfnamefont{E.~J.} \bibnamefont{Lee}},
  \bibinfo{author}{\bibfnamefont{K.~I.} \bibnamefont{Goh}},
  \bibinfo{author}{\bibfnamefont{B.}~\bibnamefont{Kahng}}, \bibnamefont{and}
  \bibinfo{author}{\bibfnamefont{D.}~\bibnamefont{Kim}},
  \bibinfo{journal}{Phys. Rev. E} \textbf{\bibinfo{volume}{71}},
  \bibinfo{pages}{056108} (\bibinfo{year}{2005}).

\bibitem[{\citenamefont{Wu et~al.}(2006)\citenamefont{Wu, Gao, Sun, and
  Huang}}]{wu2006}
\bibinfo{author}{\bibfnamefont{J.~J.} \bibnamefont{Wu}},
  \bibinfo{author}{\bibfnamefont{Z.~Y.} \bibnamefont{Gao}},
  \bibinfo{author}{\bibfnamefont{H.~J.} \bibnamefont{Sun}}, \bibnamefont{and}
  \bibinfo{author}{\bibfnamefont{H.~J.} \bibnamefont{Huang}},
  \bibinfo{journal}{Europhys. Lett.} \textbf{\bibinfo{volume}{74}},
  \bibinfo{pages}{560} (\bibinfo{year}{2006}).

\bibitem[{\citenamefont{Maslov and Sneppen}(2002)}]{maslov2002}
\bibinfo{author}{\bibfnamefont{S.}~\bibnamefont{Maslov}} \bibnamefont{and}
  \bibinfo{author}{\bibfnamefont{K.}~\bibnamefont{Sneppen}},
  \bibinfo{journal}{Science} \textbf{\bibinfo{volume}{296}},
  \bibinfo{pages}{910} (\bibinfo{year}{2002}).

\bibitem[{\citenamefont{Newman}(2003)}]{newman2003}
\bibinfo{author}{\bibfnamefont{M.~E.~J.} \bibnamefont{Newman}},
  \bibinfo{journal}{Phys. Rev. E} \textbf{\bibinfo{volume}{67}},
  \bibinfo{pages}{026126} (\bibinfo{year}{2003}).

\bibitem[{\citenamefont{V\'azquez et~al.}(2002)\citenamefont{V\'azquez,
  Pastor-Satorras, and Vespignani}}]{vazquez2002}
\bibinfo{author}{\bibfnamefont{A.}~\bibnamefont{V\'azquez}},
  \bibinfo{author}{\bibfnamefont{R.}~\bibnamefont{Pastor-Satorras}},
  \bibnamefont{and}
  \bibinfo{author}{\bibfnamefont{A.}~\bibnamefont{Vespignani}},
  \bibinfo{journal}{Phys. Rev. E} \textbf{\bibinfo{volume}{65}},
  \bibinfo{pages}{066130} (\bibinfo{year}{2002}).

\bibitem[{\citenamefont{Rual et~al.}(2005)\citenamefont{Rual, Venkatesan, Hao,
  Hirozane-Kishikawa, Dricot, Li, Berriz, Gibbons, Dreze, Ayivi-Guedehoussou
  et~al.}}]{rual2005}
\bibinfo{author}{\bibfnamefont{J.-F.} \bibnamefont{Rual}}
  \bibnamefont{et~al.}, \bibinfo{journal}{Nature (London)}
  \textbf{\bibinfo{volume}{437}}, \bibinfo{pages}{1173} (\bibinfo{year}{2005}).

\bibitem[{\citenamefont{Serrano and Bogu\~n\'a}(2006{\natexlab{c}})}]{serrano2006-1}
\bibinfo{author}{\bibfnamefont{M.~A.} \bibnamefont{Serrano}} \bibnamefont{and}
  \bibinfo{author}{\bibfnamefont{M.}~\bibnamefont{Bogu\~n\'a}},
  \bibinfo{journal}{Phys. Rev. E} \textbf{\bibinfo{volume}{74}},
  \bibinfo{pages}{056114} (\bibinfo{year}{2006}{\natexlab{c}}).

\bibitem[{\citenamefont{Guimer\`a et~al.}(2007)\citenamefont{Guimer\`a,
  Sales-Pardo, and Amaral}}]{guimera2007}
\bibinfo{author}{\bibfnamefont{R.}~\bibnamefont{Guimer\`a}},
  \bibinfo{author}{\bibfnamefont{M.}~\bibnamefont{Sales-Pardo}},
  \bibnamefont{and} \bibinfo{author}{\bibfnamefont{L.~A.~N.}
  \bibnamefont{Amaral}}, \bibinfo{journal}{Nat. Phys.}
  \textbf{\bibinfo{volume}{3}}, \bibinfo{pages}{63} (\bibinfo{year}{2007}).

\bibitem[{\citenamefont{Newman and Park}(2003)}]{newman2003-2}
\bibinfo{author}{\bibfnamefont{M.~E.~J.} \bibnamefont{Newman}}
  \bibnamefont{and} \bibinfo{author}{\bibfnamefont{J.}~\bibnamefont{Park}},
  \bibinfo{journal}{Phys. Rev. E} \textbf{\bibinfo{volume}{68}},
  \bibinfo{pages}{036122} (\bibinfo{year}{2003}).

\bibitem[{\citenamefont{Erd\H{o}s and R\'enyi}(1959)}]{erdos1959}
\bibinfo{author}{\bibfnamefont{P.}~\bibnamefont{Erd\H{o}s}} \bibnamefont{and}
  \bibinfo{author}{\bibfnamefont{A.}~\bibnamefont{R\'enyi}},
  \bibinfo{journal}{Publ. Math.-Debrecen} \textbf{\bibinfo{volume}{6}},
  \bibinfo{pages}{290} (\bibinfo{year}{1959}).

\bibitem[{\citenamefont{Barth\'elemy}(2004)}]{barthelemy2004}
\bibinfo{author}{\bibfnamefont{M.}~\bibnamefont{Barth\'elemy}},
  \bibinfo{journal}{Eur. Phys. J. B} \textbf{\bibinfo{volume}{38}},
  \bibinfo{pages}{163} (\bibinfo{year}{2004}).

\bibitem[{\citenamefont{Freeman}(1977)}]{freeman1977}
\bibinfo{author}{\bibfnamefont{L.~C.} \bibnamefont{Freeman}},
  \bibinfo{journal}{Sociometry} \textbf{\bibinfo{volume}{40}},
  \bibinfo{pages}{35} (\bibinfo{year}{1977}).

\bibitem[{\citenamefont{Goh et~al.}(2001)\citenamefont{Goh, Kahng, and
  Kim}}]{goh2001}
\bibinfo{author}{\bibfnamefont{K.~I.} \bibnamefont{Goh}},
  \bibinfo{author}{\bibfnamefont{B.}~\bibnamefont{Kahng}}, \bibnamefont{and}
  \bibinfo{author}{\bibfnamefont{D.}~\bibnamefont{Kim}},
  \bibinfo{journal}{Phys. Rev. Lett} \textbf{\bibinfo{volume}{87}},
  \bibinfo{pages}{278701} (\bibinfo{year}{2001}).

\end{thebibliography}
\end{document}